\pdfoutput=1
\documentclass[12pt,a4paper]{article}

\usepackage{ifthen} 
\newboolean{pdflatex}
\setboolean{pdflatex}{true} 

\newboolean{articletitles}
\setboolean{articletitles}{true} 

\newboolean{uprightparticles}
\setboolean{uprightparticles}{false} 

\def\paperauthors{LHCb collaboration} 
\def\paperasciititle{Measurement of the branching fraction ratio RK at large dilepton invariant mass} 
\def\papertitle{Measurement of the branching fraction ratio $R_K$ at large dilepton invariant mass} 
\def\paperkeywords{{High Energy Physics}, {LHCb}} 
\def\papercopyright{\the\year\ CERN for the benefit of the LHCb collaboration} 
\def\paperlicence{CC BY 4.0 licence}
\def\paperlicenceurl{https://creativecommons.org/licenses/by/4.0/}

\newif\ifEnableSectionTOCLinks
\EnableSectionTOCLinksfalse 

\usepackage[top=1in, bottom=1.25in, left=1in, right=1in]{geometry}

\usepackage{microtype}
\usepackage{lineno}  
\usepackage{xspace} 
\usepackage{caption} 

\usepackage{graphicx}  
\usepackage{color}
\usepackage{colortbl}
\graphicspath{{./figs/}} 

\usepackage{amsmath} 
\usepackage{amssymb}
\usepackage{amsfonts}
\usepackage{upgreek} 

\newcommand*\patchAmsMathEnvironmentForLineno[1]{%
\expandafter\let\csname old#1\expandafter\endcsname\csname #1\endcsname
\expandafter\let\csname oldend#1\expandafter\endcsname\csname
end#1\endcsname
 \renewenvironment{#1}%
   {\linenomath\csname old#1\endcsname}%
   {\csname oldend#1\endcsname\endlinenomath}%
}
\newcommand*\patchBothAmsMathEnvironmentsForLineno[1]{%
  \patchAmsMathEnvironmentForLineno{#1}%
  \patchAmsMathEnvironmentForLineno{#1*}%
}
\AtBeginDocument{%
\patchBothAmsMathEnvironmentsForLineno{equation}%
\patchBothAmsMathEnvironmentsForLineno{align}%
\patchBothAmsMathEnvironmentsForLineno{flalign}%
\patchBothAmsMathEnvironmentsForLineno{alignat}%
\patchBothAmsMathEnvironmentsForLineno{gather}%
\patchBothAmsMathEnvironmentsForLineno{multline}%
\patchBothAmsMathEnvironmentsForLineno{eqnarray}%
}

\usepackage{hyperxmp}

\usepackage[pdftex,
            pdfauthor={\paperauthors},
            pdftitle={\paperasciititle},
            pdfkeywords={\paperkeywords},
            pdfcopyright={Copyright (C) \papercopyright},
            pdflicenseurl={\paperlicenceurl}]{hyperref}

\usepackage[bottom,flushmargin,hang,multiple]{footmisc}

\usepackage[all]{hypcap} 

\usepackage{xspace} 
\usepackage{upgreek}

\def\lhcb   {\mbox{LHCb}\xspace}

\def\ecal   {ECAL\xspace}
\def\hcal   {HCAL\xspace}
\def\MagUp {\mbox{\em Mag\kern -0.05em Up}\xspace}

\ifthenelse{\boolean{uprightparticles}}%
{

 \def\Pmu         {\ensuremath{\upmu}\xspace}

 \def\Ppi         {\ensuremath{\uppi}\xspace}

 \def\Ppsi        {\ensuremath{\uppsi}\xspace}

 \def\PDelta      {\ensuremath{\Delta}\xspace}                 
 \def\PXi         {\ensuremath{\Xi}\xspace}                 
 \def\PLambda     {\ensuremath{\Lambda}\xspace}                 
 \def\PSigma      {\ensuremath{\Sigma}\xspace}                 
 \def\POmega      {\ensuremath{\Omega}\xspace}                 
 \def\PUpsilon    {\ensuremath{\Upsilon}\xspace}
 \let\oldPi\Pi
 \def\PPi         {\ensuremath{\oldPi}\xspace}

 \def\PB      {\ensuremath{\mathrm{B}}\xspace}                 
 \def\PD      {\ensuremath{\mathrm{D}}\xspace}                 
 \def\PJ      {\ensuremath{\mathrm{J}}\xspace}                 
 \def\PK      {\ensuremath{\mathrm{K}}\xspace}                 
 \def\Pb      {\ensuremath{\mathrm{b}}\xspace}                 
 \def\Pc      {\ensuremath{\mathrm{c}}\xspace}                 
                  
 \def\Pe      {\ensuremath{\mathrm{e}}\xspace}                 
 \def\Ps      {\ensuremath{\mathrm{s}}\xspace}

 \def\thebaroffset{0.0em}
}
{

 \def\Pmu         {\ensuremath{\mu}\xspace}

 \def\Ppi         {\ensuremath{\pi}\xspace}

 \def\Ppsi        {\ensuremath{\psi}\xspace}                 
                  
 \mathchardef\PDelta="7101
 \mathchardef\PXi="7104
 \mathchardef\PLambda="7103
 \mathchardef\PSigma="7106
 \mathchardef\POmega="710A
 \mathchardef\PUpsilon="7107
 \mathchardef\PPi="7105
 \def\PB      {\ensuremath{B}\xspace}                 
 \def\PD      {\ensuremath{D}\xspace}                 
 \def\PJ      {\ensuremath{J}\xspace}                 
 \def\PK      {\ensuremath{K}\xspace}                 
 \def\Pb      {\ensuremath{b}\xspace}                 
 \def\Pc      {\ensuremath{c}\xspace}                 
                  
 \def\Pe      {\ensuremath{e}\xspace}                 
 \def\Ps      {\ensuremath{s}\xspace}

 \def\thebaroffset{0.18em}
}
\newcommand{\offsetoverline}[2][\thebaroffset]{\kern #1\overline{\kern -#1 #2}}%

\makeatletter
\ifcase \@ptsize \relax
  \newcommand{\miniscule}{\@setfontsize\miniscule{4}{5}}
\or
  \newcommand{\miniscule}{\@setfontsize\miniscule{5}{6}}
\or
  \newcommand{\miniscule}{\@setfontsize\miniscule{5}{6}}
\fi
\makeatother

\DeclareRobustCommand{\optbar}[1]{\shortstack{{\miniscule (\rule[.5ex]{1.25em}{.18mm})}
  \\ [-.7ex] $#1$}}

\def\electron   {{\ensuremath{\Pe}}\xspace}
\def\en         {{\ensuremath{\Pe^-}}\xspace}   
\def\ep         {{\ensuremath{\Pe^+}}\xspace}

\def\epem       {{\ensuremath{\Pe^+\Pe^-}}\xspace}

\def\muon       {{\ensuremath{\Pmu}}\xspace}
\def\mup        {{\ensuremath{\Pmu^+}}\xspace}
\def\mun        {{\ensuremath{\Pmu^-}}\xspace} 

\def\mumu       {{\ensuremath{\Pmu^+\Pmu^-}}\xspace}

\def\lepton     {{\ensuremath{\ell}}\xspace}
\def\ellm       {{\ensuremath{\ell^-}}\xspace}
\def\ellp       {{\ensuremath{\ell^+}}\xspace}

\def\ellell     {\ensuremath{\ell^+ \ell^-}\xspace}

\def\squark    {{\ensuremath{\Ps}}\xspace}

\def\cquark    {{\ensuremath{\Pc}}\xspace}

\def\bquark    {{\ensuremath{\Pb}}\xspace}

\def\pion   {{\ensuremath{\Ppi}}\xspace}
\def\piz    {{\ensuremath{\pion^0}}\xspace}
\def\pip    {{\ensuremath{\pion^+}}\xspace}
\def\pim    {{\ensuremath{\pion^-}}\xspace}

\def\kaon    {{\ensuremath{\PK}}\xspace}

\def\KorKbar {\kern \thebaroffset\optbar{\kern -\thebaroffset \PK}{}\xspace}

\def\Kp      {{\ensuremath{\kaon^+}}\xspace}
\def\Km      {{\ensuremath{\kaon^-}}\xspace}

\def\Kstar   {{\ensuremath{\kaon^*}}\xspace}

\def\D       {{\ensuremath{\PD}}\xspace}

\def\DorDbar {\kern \thebaroffset\optbar{\kern -\thebaroffset \PD}\xspace}
\def\Dz      {{\ensuremath{\D^0}}\xspace}

\def\Dp      {{\ensuremath{\D^+}}\xspace}
\def\Dm      {{\ensuremath{\D^-}}\xspace}

\def\DpDm    {\ensuremath{\Dp {\kern -0.16em \Dm}}\xspace}

\def\Dstarp  {{\ensuremath{\D^{*+}}}\xspace}

\def\B       {{\ensuremath{\PB}}\xspace}

\def\BorBbar {\kern \thebaroffset\optbar{\kern -\thebaroffset \PB}\xspace}
\def\Bz      {{\ensuremath{\B^0}}\xspace}

\def\Bd      {{\ensuremath{\B^0}}\xspace}

\def\BdorBdbar {\kern \thebaroffset\optbar{\kern -\thebaroffset \Bd}\xspace}
\def\Bu      {{\ensuremath{\B^+}}\xspace}

\def\Bp      {{\ensuremath{\Bu}}\xspace}

\def\Bs      {{\ensuremath{\B^0_\squark}}\xspace}

\def\BsorBsbar {\kern \thebaroffset\optbar{\kern -\thebaroffset \Bs}\xspace}

\def\jpsi     {{\ensuremath{{\PJ\mskip -3mu/\mskip -2mu\Ppsi}}}\xspace}
\def\psitwos  {{\ensuremath{\Ppsi{(2S)}}}\xspace}

\def\Y#1S{\ensuremath{\PUpsilon{(#1S)}}\xspace}

\def\LorLbar     {\kern \thebaroffset\optbar{\kern -\thebaroffset \PLambda}\xspace}

\def\BF         {{\ensuremath{\mathcal{B}}}\xspace}
\def\BR         {\BF}

\newcommand{\decay}[2]{\mbox{\ensuremath{#1\!\to #2}}\xspace} 

\def\to                 {\ensuremath{\rightarrow}\xspace}

\def\qsq       {{\ensuremath{q^2}}\xspace}

\def\bsll     {\decay{\bquark}{\squark \ell^+ \ell^-}}

\def\AT#1     {\ensuremath{A_{\mathrm{T}}^{#1}}\xspace}           

\def\C#1      {\ensuremath{\mathcal{C}_{#1}}\xspace}                       
\def\Cp#1     {\ensuremath{\mathcal{C}_{#1}^{'}}\xspace}                    
\def\Ceff#1   {\ensuremath{\mathcal{C}_{#1}^{\mathrm{(eff)}}}\xspace}        
\def\Cpeff#1  {\ensuremath{\mathcal{C}_{#1}^{'\mathrm{(eff)}}}\xspace}       
\def\Ope#1    {\ensuremath{\mathcal{O}_{#1}}\xspace}                       
\def\Opep#1   {\ensuremath{\mathcal{O}_{#1}^{'}}\xspace}                    

\newcommand{\nospaceunit}[1]{\ensuremath{\text{#1}}}       
\newcommand{\aunit}[1]{\ensuremath{\text{\,#1}}}       

\newcommand{\tev}{\aunit{Te\kern -0.1em V}\xspace}
\newcommand{\gev}{\aunit{Ge\kern -0.1em V}\xspace}
\newcommand{\mev}{\aunit{Me\kern -0.1em V}\xspace}
\newcommand{\kev}{\aunit{ke\kern -0.1em V}\xspace}
\newcommand{\ev}{\aunit{e\kern -0.1em V}\xspace}
 
\newcommand{\mevc}{\ensuremath{\aunit{Me\kern -0.1em V\!/}c}\xspace}
\newcommand{\gevc}{\ensuremath{\aunit{Ge\kern -0.1em V\!/}c}\xspace}
\newcommand{\mevcc}{\ensuremath{\aunit{Me\kern -0.1em V\!/}c^2}\xspace}
\newcommand{\gevcc}{\ensuremath{\aunit{Ge\kern -0.1em V\!/}c^2}\xspace}
\newcommand{\gevgevcccc}{\ensuremath{\gev^2\!/c^4}\xspace} 

\def\mum  {\ensuremath{\,\upmu\nospaceunit{m}}\xspace}

\def\fb   {\ensuremath{\aunit{fb}}\xspace}
\def\invfb   {\ensuremath{\fb^{-1}}\xspace}

\newcommand{\stat}{\aunit{(stat)}\xspace}
\newcommand{\syst}{\aunit{(syst)}\xspace}

\newcommand{\chisq}{\ensuremath{\chi^2}\xspace}

\def\deriv {\ensuremath{\mathrm{d}}}

\def\gsim{{~\raise.15em\hbox{$>$}\kern-.85em
          \lower.35em\hbox{$\sim$}~}\xspace}
\def\lsim{{~\raise.15em\hbox{$<$}\kern-.85em
          \lower.35em\hbox{$\sim$}~}\xspace}

\def\sPlot{\mbox{\em sPlot}\xspace}

\def\pt         {\ensuremath{p_{\mathrm{T}}}\xspace}

\def\ptot       {\ensuremath{p}\xspace}
\def\et         {\ensuremath{E_{\mathrm{T}}}\xspace}

\def\evtgen     {\mbox{\textsc{EvtGen}}\xspace}

\def\geant      {\mbox{\textsc{Geant4}}\xspace}

\def\photos     {\mbox{\textsc{Photos}}\xspace}

\def\pythia     {\mbox{\textsc{Pythia}}\xspace}

\def\tell1  {TELL1\xspace}
\def\ukl1   {UKL1\xspace}

\newcommand{\lhcborcid}[1]{\href{https://orcid.org/#1}{\hspace*{0.1em}\raisebox{-0.45ex}{\includegraphics[width=1em]{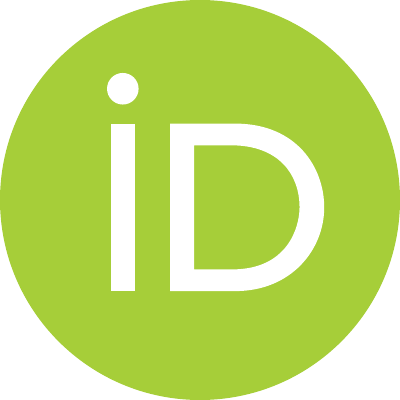}}}}

\newcommand{\BKll}{\mbox{\decay{\Bp }{ \Kp \ellp \ellm}}\xspace}
\newcommand{\BKmumu}{\mbox{\decay{\Bp}{\Kp \mumu}}\xspace}
\newcommand{\BKee}{\mbox{\decay{\Bp }{ \Kp \ep \en}}\xspace}
\newcommand{\BKemu}{\mbox{\decay{\Bp }{ \Kp \ep \mu^-}}\xspace}
\newcommand{\Kll}{\ensuremath{\Kp \ellell}\xspace}
\newcommand{\Kee}{\ensuremath{\Kp \ep \en}\xspace}
\newcommand{\Kmumu}{\ensuremath{\Kp\mumu}\xspace}

\newcommand{\BKJPsill}{\mbox{\decay{\Bp }{ \Kp \jpsi(\to \ellell)}}\xspace}
\newcommand{\BKJPsiee}{\mbox{\decay{\Bp }{ \Kp \jpsi(\to \epem)}}\xspace}
\newcommand{\BKJPsimumu}{\mbox{\decay{\Bp }{ \Kp \jpsi(\to \mumu)}}\xspace}

\newcommand{\BKpsiSll}{\mbox{\decay{\Bp }{ \Kp \psitwos(\to \ellell)}}\xspace}
\newcommand{\BKpsiSee}{\mbox{\decay{\Bp }{ \Kp \psitwos(\to \ep \en)}}\xspace}
\newcommand{\BKpsiSmumu}{\mbox{\decay{\Bp }{ \Kp \psitwos(\to \mumu)}}\xspace}

\newcommand{\BPipsiSee}{\mbox{\decay{\Bp }{ \pip \psitwos(\to \ep \en)}}\xspace}

\newcommand{\BKpipill}{\mbox{\decay{\Bp }{ \Kp \pip \pim}}\xspace}
\newcommand{\BKpipinomisID}{\mbox{\decay{\Bp }{ \Kp \pip \pim}}\xspace}
\newcommand{\BKKKnomisID}{\mbox{\decay{\Bp }{ \Kp \Kp \Km}}\xspace}

\newcommand{\RK}{{\ensuremath{R_K}}\xspace}
\newcommand{\rJPsi}{{\ensuremath{r_{\jpsi}}}\xspace}
\newcommand{\RPsiS}{{\ensuremath{R_{\psitwos}}}\xspace}

\newcommand{\qsqmin}{\ensuremath{q^{\rm{2}}_{\rm{min}}}\xspace}
\newcommand{\qsqmax}{\ensuremath{q^{\rm{2}}_{\rm{max}}}\xspace}

\newcommand{\qsqTrack}{\ensuremath{q^{\rm{2}}_{\rm{track}}}\xspace}
\newcommand{\qsqTrue}{\ensuremath{q^{\rm{2}}_{\rm{true}}}\xspace}
\newcommand{\mkee}{\ensuremath{{m(\Kee)}}\xspace}
\newcommand{\mkmumu}{\ensuremath{{m(\Kmumu)}}\xspace}

\newcommand{\mDTFJPsi}{\ensuremath{m_{\jpsi}}\xspace}
\newcommand{\mDTFPsitwos}{\ensuremath{m_{\psitwos}}\xspace}

\newcommand{\mkeeconst}{\ensuremath{{\mDTFJPsi(\Kee)}}\xspace}
\newcommand{\mkmumuconst}{\ensuremath{{\mDTFJPsi(\Kmumu)}}\xspace}
\newcommand{\mkllconst}{\ensuremath{{\mDTFJPsi(\Kll)}}\xspace}

\newcommand{\mkmumuconstpsitwos}{\ensuremath{{\mDTFPsitwos(\Kmumu)}}\xspace}
\newcommand{\mkllconstpsitwos}{\ensuremath{{\mDTFPsitwos(\Kll)}}\xspace}

\newcommand{\chiSqBVtx}{{\ensuremath{\chi^2_{\rm{DV}}(\Bp)}}\xspace}
\newcommand{\ptransverse}{{\ensuremath{p_{\rm{T}}}}\xspace}
\newcommand{\ipChiSq}{\ensuremath{\chi^2_{\rm{IP}}}\xspace}

\newcommand\ddfrac[2]{\frac{\displaystyle #1}{\displaystyle #2}}

\newcommand{\figref}[1]{Fig.~\ref{#1}}
\newcommand{\tabref}[1]{Table~\ref{#1}}

\newcommand{\secref}[1]{Sec.~\ref{#1}}
\newcommand{\equref}[1]{Eq.~\ref{#1}}

\newcommand{\dif}{\ensuremath{\textrm{d}}\xspace}
\newcommand{\qsqhightrack}{{\ensuremath{\qsqTrack>14.3\gevgevcccc}}\xspace}

\hypersetup{
  colorlinks   = true, 
  urlcolor     = blue, 
  linkcolor    = blue, 
  citecolor    = red   
}

\ifEnableSectionTOCLinks
    \usepackage[explicit]{titlesec} 
    
    \let\oldcontentsline\contentsline
    \renewcommand

    \titleformat{\section}{\normalfont\Large\bf}{\hyperlink{tocsection.\thesection}{{\thesection} \parbox[t]{\dimexpr\textwidth-1pc}{#1}}}{1pc}{}

    \titleformat{\subsection}{\normalfont\bf}{\hyperlink{tocsubsection.\thesubsection}{{\thesubsection} \parbox[t]{\dimexpr\textwidth-1pc}{#1}}}{1pc}{}

    \titleformat{name=\section,numberless}[display]{}{}{0pt}{\normalfont\Huge\bfseries #1}
\fi

\usepackage{cite} 
\usepackage{mciteplus}

\usepackage{longtable} 
\usepackage{booktabs}

\begin{document}

\renewcommand{\thefootnote}{\fnsymbol{footnote}}
\setcounter{footnote}{1}


\begin{titlepage}
\pagenumbering{roman}

\vspace*{-1.5cm}
\centerline{\large EUROPEAN ORGANIZATION FOR NUCLEAR RESEARCH (CERN)}
\vspace*{1.5cm}
\noindent
\begin{tabular*}{\linewidth}{lc@{\extracolsep{\fill}}r@{\extracolsep{0pt}}}
\ifthenelse{\boolean{pdflatex}}
{\vspace*{-1.5cm}\mbox{\!\!\!\includegraphics[width=.14\textwidth]{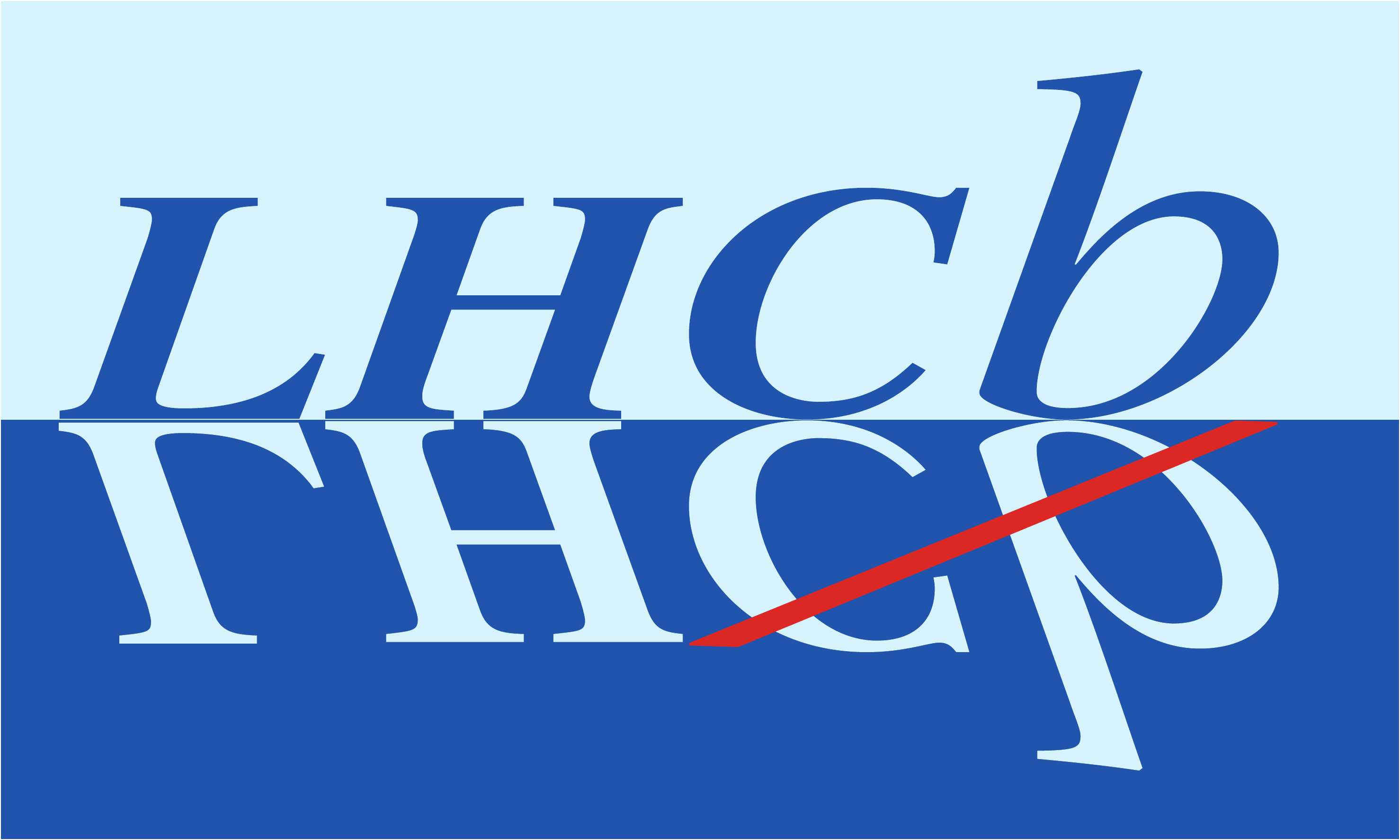}} & &}%
{\vspace*{-1.2cm}\mbox{\!\!\!\includegraphics[width=.12\textwidth]{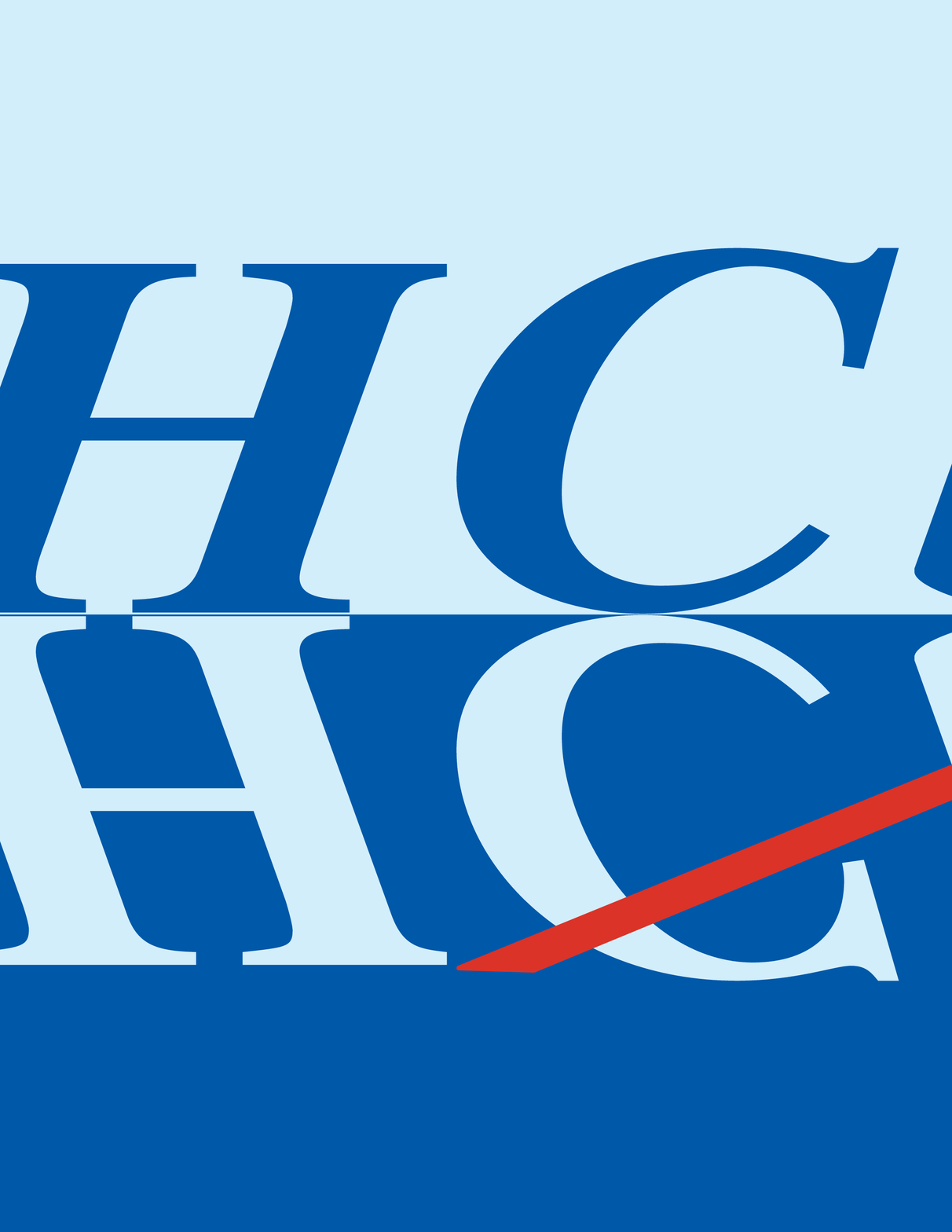}} & &}%
\\
 & & CERN-EP-2025-069 \\  
 & & LHCb-PAPER-2024-056 \\  
 & & May 6, 2025 \\ 
 & & \\
\end{tabular*}

\vspace*{4.0cm}

{\normalfont\bfseries\boldmath\huge
\begin{center}
  \papertitle 
\end{center}
}

\vspace*{2.0cm}

\begin{center}
\paperauthors\footnote{Authors are listed at the end of this paper.}
\end{center}

\vspace{\fill}

\begin{abstract}
  \noindent A test of lepton universality between muons and electrons is performed using \decay{\Bu}{\Kp \ellp \ellm} decays (where $\lepton = \electron$, $\muon$), in the dilepton invariant-mass-squared region above 14.3\gevgevcccc. The data used for the measurement consists of beauty meson decays produced in proton-proton collisions, corresponding to an integrated luminosity of $9\invfb$, collected by the \mbox{LHCb} experiment between 2011 and 2018. The ratio of branching fractions for \decay{\Bu}{\Kp \mup \mun}  and \decay{\Bu}{\Kp \ep \en} decays is measured to be  \mbox{$R_K = 1.08^{+0.11}_{-0.09}\;(\text{stat})\;^{+0.04}_{-0.04}\;(\text{syst})$}, which is consistent with the Standard Model prediction of unity. This constitutes the most precise test of lepton flavour universality using \decay{\Bu}{\Kp \ellp \ellm}  decays with dilepton invariant-mass-squared above the \psitwos mass, whilst being the first of its kind at a hadron collider. 
  
\end{abstract}

\vspace*{2.0cm}

\begin{center}
  Published in JHEP 07 (2025) 198 
\end{center}

\vspace{\fill}

{\footnotesize 
\centerline{\copyright~\papercopyright. \href{\paperlicenceurl}{\paperlicence}.}}
\vspace*{2mm}

\end{titlepage}


\newpage
\setcounter{page}{2}
\mbox{~}
%
%
%
%


\renewcommand{\thefootnote}{\arabic{footnote}}
\setcounter{footnote}{0}


\cleardoublepage


\pagestyle{plain} 
\setcounter{page}{1}
\pagenumbering{arabic}


\section{Introduction}
\label{sec:Introduction}
The decay of a \bquark quark to an \squark quark and two oppositely charged leptons (\bsll) is an example of a flavour-changing neutral current transition. In the Standard Model~(SM), such processes are loop-induced because no SM gauge boson mediates the decay at tree level. Processes involving a \bsll transition, such as \decay{\B}{\kaon^{(*)}\ellell} and \decay{\Bs}{ \phi\mumu} decays,\footnote{In this paper, the inclusion of the charge-conjugate mode is implied.} are therefore exceedingly rare with branching fractions of order $\mathcal{O}(10^{-7})$. Extensions to the SM often hypothesise the existence of new particles that are not directly detectable with current experimental resources but can contribute sizeable modifications to the properties of \bsll transitions~\cite{Celis:2015ara,Falkowski:2015zwa,Crivellin:2015mga,Barbieri:2016las,Buttazzo:2017ixm}. Discrepancies with respect to the SM prediction ranging from $2\sigma$ to $4\sigma$ have been observed in the measurements of branching fractions and angular observables for the decays \decay{\Bs}{ \phi\mumu}, \decay{\Bp}{ \Kp\mumu} and \decay{\Bz}{ \Kstar(892)^0\mumu}~\cite{LHCb-PAPER-2024-011,LHCb-PAPER-2023-033,LHCb-PAPER-2023-032,LHCb-PAPER-2021-022,LHCb-PAPER-2021-014,CMS:2024atz}. 

Calculating SM predictions for \bsll transitions requires an understanding of hadronic form factors~\cite{Bharucha_2016,Horgan_2014} and nonperturbative long-distance effects~\cite{Gubernari_2021}, both of which are known imprecisely, complicating the interpretation of the discrepancies.
However, observables involving ratios of \bsll branching fractions for two different lepton flavours have minimal hadronic theory uncertainty, as they cancel out because the strong force does not couple to leptons. Furthermore, the Yukawa couplings for the leptons are small compared to their SM gauge couplings, leading to an accidental symmetry known as lepton universality (LU). This principle has been tested across various processes~\cite{Schael:1516169,PhysRevLett.115.071801,Lazzeroni:1434415,KEDR:2013dpd}, and is most precisely confirmed in Z boson decays at the per mille level~\cite{ALEPH:2005ab}. Consequently, ratios of branching fractions involving \bsll transitions only deviate from unity due to mass-related phase-space and QED effects, which are small for ratios involving light leptons. In particular, QED effects are below $1\%$ at dilepton invariant mass squared (\qsq) lower than $6 \gevgevcccc$~\cite{Bordone:2016gaq} and up to $4\%$ at \qsq above the \psitwos resonance. These effects have been shown to be correctly modelled in the whole accessible phase space by the simulation software employed~\cite{Bordone:2016gaq,Isidori_2022}. Measuring ratios of branching fractions is further motivated by an interest in beyond Standard Model theories that introduce LU-violating extensions~\cite{PhysRevD.69.074020,DORSNER20161,PhysRevD.89.095033}.

Decays involving electrons or muons in the final state, rather than taus, are most commonly studied due to their experimental accessibility. Tests of LU have been conducted by the BaBar~\cite{PhysRevD.86.032012}, Belle~\cite{belle_rk}, CMS~\cite{CMS_Collaboration_2024} and LHCb collaborations~\cite{LHCb-PAPER-2021-038, LHCb-PAPER-2022-045,LHCb-PAPER-2022-046,LHCb-PAPER-2024-032,LHCb-PAPER-2019-040,LHCb-PAPER-2024-046} using a wide variety of decay channels involving \bsll transitions, yielding results that are consistent with SM expectation.
The most precise LU measurements involving \BKll decays have been performed in the \qsq region below the \jpsi resonance, with a precision of 4\%~\cite{LHCb-PAPER-2022-045,LHCb-PAPER-2022-046}, while the only existing LU measurement of \BKll in the high-\qsq region is that performed by the Belle collaboration~\cite{belle_rk}.
Measurements in the high-\qsq region, above the \psitwos resonance, are complementary to low-\qsq measurements having different background and efficiency considerations.

This paper describes a test of LU in the decays of \decay{\Bp}{ \Kp\mumu} and \BKee performed by measuring the branching fraction ratio
\begin{equation}
    \RK \equiv \ddfrac{\int_{\qsqmin}^{\qsqmax} \ddfrac{\dif \BR(\BKmumu)}{\dif\qsq} \dif\qsq}{\int_{\qsqmin}^{\qsqmax} \ddfrac{\dif \BR(\BKee)}{\dif\qsq}\dif\qsq},
\end{equation}
where the integral limits define the \qsq bin of the measurement, which for this analysis corresponds to $\qsq > 14.3\gevgevcccc$. The measurement is performed using a dataset corresponding to 9\invfb of integrated luminosity collected by the LHCb experiment between 2011 and 2018. Similar to other LU ratios in this system, the measurement is performed relative to the normalisation mode \BKJPsill in order to cancel and control many systematic uncertainties associated with the computation of efficiencies.  The measured yield of \BKee decays was kept blind until the analysis was finalised to avoid experimenter's bias.

This paper is structured as follows: the \lhcb detector and simulation are outlined in \secref{sec:detector_and_simulation}. The selection requirements used to isolate signal decays, including details on how backgrounds such as \BKpsiSee leakage are suppressed to negligible levels, are presented in \secref{sec:selection}. In \secref{sec:eff}, the corrections to the simulation samples and associated cross-checks are detailed. A novel method of extracting \RK that significantly reduces systematic uncertainties related to the imprecisely understood shape of the high-\qsq spectrum is presented in \secref{sec:fits}, along with the determination of the rare \BKll muon and electron channel yields. Systematic uncertainties are discussed in \secref{sec:systs}, and the result is presented in \secref{sec:results}.

\section{Detector and simulation}\label{sec:detector_and_simulation}
The \lhcb detector~\cite{LHCb-DP-2008-001,LHCb-DP-2014-002} is a single-arm forward spectrometer covering the \mbox{pseudorapidity} range $2<\eta <5$, designed for the study of particles containing \bquark or \cquark quarks.
The detector includes a high-precision tracking system consisting of a silicon-strip vertex detector surrounding the $pp$ interaction region~\cite{LHCb-DP-2014-001}, a large-area silicon-strip detector located upstream of a dipole magnet with a bending power of about $4{\mathrm{\,T\,m}}$, and three stations of silicon-strip detectors and straw drift tubes~\cite{LHCb-DP-2013-003,LHCb-DP-2017-001} placed downstream of the magnet.
The tracking system provides a measurement of the momentum, \ptot, of charged particles with a relative uncertainty that varies from 0.5\% at low momentum to 1.0\% at 200\gevc.
The minimum distance of a track to a primary $pp$ collision vertex (PV), the impact parameter, is measured with a resolution of $(15+29/\pt)\mum$, where \pt is the component of the momentum transverse to the beam, in\,\gevc.
Different types of charged hadrons are distinguished using information from two ring-imaging Cherenkov detectors (RICH)~\cite{LHCb-DP-2012-003}.
 Photons, electrons and hadrons are identified by a calorimeter system consisting of scintillating-pad and preshower detectors, as well as an electromagnetic calorimeter (\ecal) and a hadronic calorimeter (\hcal).
 Muons are identified by a system composed of alternating layers of iron and multiwire proportional chambers~\cite{LHCb-DP-2012-002}.
The online event selection is performed by a trigger~\cite{LHCb-DP-2012-004},  which consists of a hardware stage, based on information from the calorimeter and muon systems, followed by a software stage, which applies a full event reconstruction.

Simulated events are used to optimise the signal selection, model the signal and backgrounds, and calculate the relative efficiency between the signal and normalisation channels. In the simulation, $pp$ collisions are generated using \pythia~\cite{Sjostrand:2006za,*Sjostrand:2007gs} with a specific \lhcb configuration~\cite{LHCb-PROC-2010-056}. Decays of hadronic particles are described by \evtgen~\cite{Lange:2001uf}, in which final-state radiation is generated using \photos~\cite{Golonka:2005pn}. The interaction of the generated particles with the detector and its response are implemented using the \geant toolkit~\cite{Allison:2006ve, *Agostinelli:2002hh} as described in Ref.~\cite{LHCb-PROC-2011-006}.

\section{Selection}\label{sec:selection}
Selection requirements are applied to the data in order to isolate the signal and reduce background contamination using both the ROOT framework~\cite{ROOT,roofit} and the Scikit-HEP ecosystem~\cite{py:scikit-hep:2020,py:uproot,py:mplhep,py:hist,py:boost-histogram,py:particle} including \textsc{zfit} ~\cite{py:zfit,py:zfit:2020}.
Background contributions fall into two broad categories: combinatorial candidates, whereby the combined tracks do not originate from a common particle decay, and decays where tracks, despite originating from a common source, are incorrectly reconstructed by missing and/or misidentifying particles. 

Since both muon and electron final states consist of three tracks originating from the same decay vertex, there is significant overlap in the requirements applied to their topological and geometric properties. In the following, the areas where the selection requirements differ between the muon and electron data are described.

Events are selected based on the decision of a hardware trigger algorithm. For the muon mode candidates, a high-\ptransverse muon is required as measured using information from the muon stations. For the electron mode candidates, an electron with high transverse energy is required, defined as $\et=E\sin{\theta}$, where $E$ is the measured energy in a cluster of cells in the \ecal, and $\theta$ is the angle between the beam direction and the line connecting the PV to the \ecal cell cluster.
In a subsequent trigger stage, final-state particles satisfying minimum \ptransverse and impact parameter requirements are then combined to construct \BKll candidates. Each candidate is required to form a good-quality vertex with a significant displacement from the primary $pp$ collisions.
To reduce contamination from misidentified backgrounds such as \BKpipill decays, a tight selection is made on particle identification (PID) variables constructed using information from all relevant subdetectors.

\begin{figure}[tb]
  \begin{center}
    \centering
    \includegraphics[page=1]{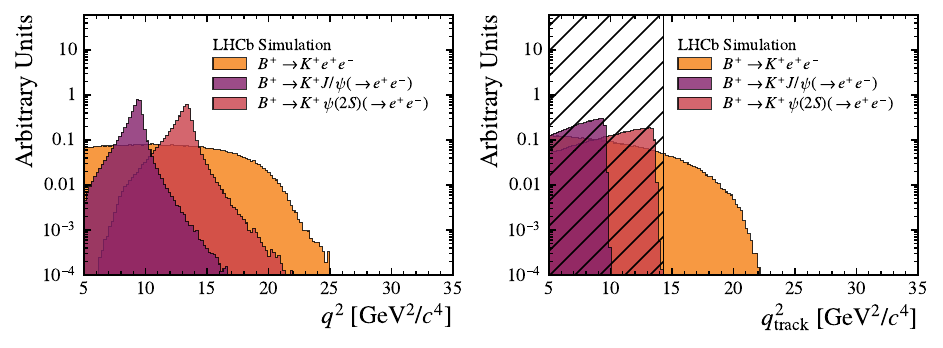}
    \vspace*{-0.5cm}
  \end{center}
  \caption{
    \small Normalised distributions of (left) \qsq and (right) \qsqTrack for simulated nonresonant \BKee signal, and the resonant decays \BKJPsiee and \BKpsiSee. The hatched region illustrates the impact of the $\qsqTrack$ selection.}
  \label{fig:q2_track_selection}
\end{figure}

While the identification of kaon and muon candidates employs information from the RICH detectors, the \hcal and the muon chambers, the identification of electrons relies principally on the measured ratio of energy deposited in showers in the ECAL and the reconstructed momenta of associated tracks~\cite{LHCb-TDR-002}. The significantly higher likelihood for electrons to emit bremsstrahlung radiation whilst traversing the detector strongly affects their momentum resolution. To mitigate this effect, a recovery algorithm~\cite{Deschamps:691634} searches the extrapolated track position of electrons in the ECAL for photon energy deposits not associated with any other charged tracks. Energy deposits with $\et > 75\mev$ are added to the electron's momentum, and the reconstructed dilepton invariant mass is computed with these photon contributions included. However, incorrectly assigning bremsstrahlung deposits to electrons from \BKJPsiee and \BKpsiSee decays results in candidates from these channels leaking upwards into the nonresonant high-\qsq region, as depicted on the left-hand side of \figref{fig:q2_track_selection}. Using the requirement $\qsq > 14.3 \gevgevcccc$ to isolate the high-\qsq region would result in an unmanageable \BKpsiSee background. However, the selection requirement \qsqhightrack, where \qsqTrack is computed using momenta estimated by the tracking system only, reduces the background from \BKpsiSee decays to negligible levels, as illustrated on the right-hand side of \figref{fig:q2_track_selection}. Selecting candidates with the \qsqTrack variable, rather than \qsq, approximately halves the expected number of rare \BKee candidates. Despite this, there is an improvement in the \RK precision in this higher purity selection, and hence using \qsqTrack is the adopted selection strategy. To ensure efficiency-related systematic uncertainties associated with modelling the resolution in the simulation cancel out, the normalisation is performed relative to the \BKJPsiee control mode, which is also selected using the \qsqTrack variable. Radiative bremsstrahlung losses for muons are insignificant compared to electrons. Therefore, the bremsstrahlung recovery procedure is not used for the reconstruction of muonic final states.

\begin{figure}[tb]
  \begin{center}
    \centering
    \includegraphics[page=1]{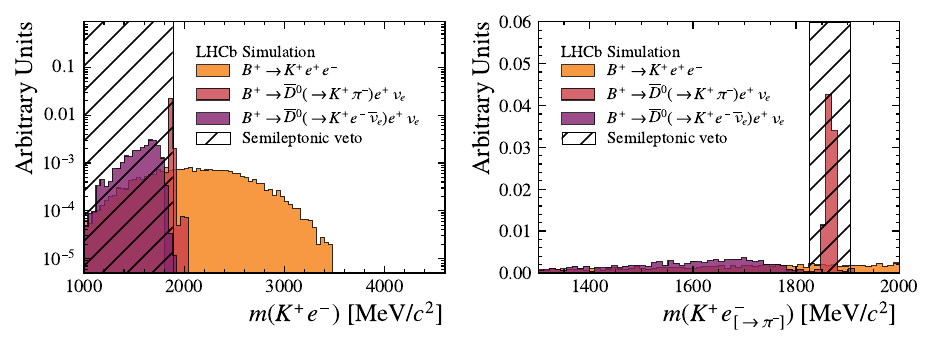}
    \vspace*{-0.5cm}
  \end{center}
  \caption{
    \small Normalised distributions of (left) $m(\Kp\en)$ and (right) $m(\Kp\en_{[\to\pim]})$ for simulated \BKee decays and the most significant semileptonic decay background channels. The regions that are vetoed are indicated by the hatched region.}
  \label{fig:cascade_cut}
\end{figure}

Semileptonic \decay{\bquark}{\cquark} decays followed by the subsequent decay of the intermediate charm hadron form a significant source of partially reconstructed background as such channels have branching fractions orders of magnitude higher than the signal.
There exists a plethora of such channels which can be split into two classes depending on whether any final-state particles have been misidentified. Both classes are efficiently removed with two selection criteria, $m(\Kp\ellm)>1885\mevcc$ and a $\PD$-mass window cut of $m(\Kp\ellm_{[\to\pim]}) \notin m(\Dz)\pm40\mevcc$, where  $m(\Dz)$ denotes the known \Dz mass~\cite{PDG2024}. The variable $m(\Kp\ellm_{[\to\pim]})$ is computed by assigning the pion mass hypothesis to the lepton and ignoring contributions from the bremsstrahlung recovery algorithm. The impact of these cuts on signal and background simulation samples in the electron mode is shown in \figref{fig:cascade_cut}. 

Boosted Decision Tree (BDT)~\cite{chen2016xgboost} classifiers are trained to distinguish \BKll signal decays from combinatorial background. The classification is achieved by exploiting differences in the distributions of various vertex-quality and kinematic variables between signal and combinatorial candidates. Additionally, quantities that relate to how isolated the candidate is from other particles in the event are utilised. The samples used to train the BDT classifier include simulation samples to represent signal decays and selected candidates in the data with $m(\Kp\ellp\ellm) > 5400\mevcc$ to represent the combinatorial background. Separate BDT classifiers are trained for the muon and electron channels.

In the electron channel, a further BDT classifier is trained to reject partially reconstructed decays of the type  ${\decay{ \Bz }{ \Kstar(892)^0(\to \Kp\pim)\epem}}$ and ${\decay{ \Bu } { \Kstar(892)^+(\to \Kp\piz)\epem}}$, where the pion is not reconstructed. The BDT classifier is trained with features similar to the combinatorial BDT classifier; however, the background is represented by simulated partially reconstructed decays. This classifier is most effective at suppressing partially reconstructed candidates with missing charged particles since charged particles produce a clearer signature in the \lhcb detector than neutral particles.

Requirements on the output of these BDT classifiers are optimised simultaneously by employing pseudoexperiments (see \secref{sec:fits}). The pair of requirements that yields the minimum average uncertainty on \RK, accounting for the statistical and the dominant systematic contributions, across the pseudoexperiments is then used. 
The requirement on the combinatorial BDT output for the muon mode is determined by optimising the expected signal significance. For this calculation, the signal yield is estimated using measured values of the \BKmumu and \BKJPsimumu branching fractions~\cite{LHCb-PAPER-2016-045, PDG2024}. The background yield is determined by extrapolating the combinatorial yield observed in the upper invariant-mass sideband into the region under the signal.

The \Bp candidates are required to have a three-body invariant mass, $m(\Kp\ellp\ellm)$, around the known \Bp mass~\cite{PDG2024}. The invariant-mass ranges used for the electron and muon channels differ significantly due to their drastically different resolution. Candidates in the rare muon channel are required to have $5180<m(\Kp\mup\mun)<5600\mevcc$, which is chosen to completely exclude partially reconstructed decays in the lower invariant-mass sideband. On the other hand, candidates in the rare electron channel are required to have \mbox{$4300<m(\Kee)<6300\mevcc$}. The wider invariant-mass requirement improves the precision with which the rare electron yield can be determined, as discussed further in \secref{sec:electron_fit}.

The selection applied to the normalisation channels is identical to the rare channels, with the exception of the $\qsq$ and $m(\Kp\ellp\ellm)$ ranges. Due to the good \qsq resolution for muonic decays, the requirement of $8.68<\qsq<10.09\gevgevcccc$ is applied to the \BKJPsimumu mode. The \qsqTrack selection applied to the \BKJPsiee channel is chosen to be that which best aligns the \mkee distribution between the resonant \BKJPsiee and nonresonant \BKee simulation. A selection of $7.1<\qsqTrack<10.0\gevgevcccc$ is employed to ensure the maximal amount of kinematic overlap between the normalisation channel and the rare signal channel.

For the normalisation channels, improvement in the resolution of the three-body invariant mass can be achieved by constraining the invariant mass of the dilepton system to that of the relevant intermediate resonance. This is implemented via the \emph{Decay Tree Fitter}~\cite{Hulsbergen:2005pu} algorithms. The invariant-mass fit of the \BKJPsimumu channel is performed in the range $5180<\mkmumuconst<5600\mevcc$ whereas the slightly wider range $5080<\mkeeconst<5680\mevcc$ is used for the \BKJPsiee channel. The efficiency of the $m(\Kp\ellp\ellm)$ selection requirement in all channels is above $99\%$. Therefore, any efficiency-related systematic uncertainty associated with differences in the invariant-mass selection applied to the normalisation and rare channels is negligible.

Finally, for events with more than one candidate after the full selection, the candidate with the highest combinatorial BDT response is retained. The fraction of rejected multiple candidates is at the subpercent level.

\section{Efficiency correction and cross-checks}\label{sec:eff}
The efficiencies are computed from simulation that is corrected using data following the procedure described in Refs.~\cite{LHCb-PAPER-2021-004,LHCb-PAPER-2023-038}. Each correction is applied with respect to the preceding one. Firstly, the tracking efficiency for electrons is determined using a tag-and-probe method applied to a sample of $\BKJPsiee$ decays~\cite{LHCb-DP-2019-003}. The difference in tracking efficiency between data and simulated samples is corrected as a function of the pseudorapidity, the azimuthal angle $\phi$ and the \pt of the probe particle. The effect of these corrections is less than 1\%, reflecting the good agreement between simulation and data. The efficiency of the PID requirements is obtained using data control channels that can be efficiently selected without the use of PID criteria~\cite{LHCb-PUB-2016-021}.

The samples include ${\Dstarp\to \Dz(\to \Km\pip) \pip}$ decays for hadron (mis)identification, ${\BKJPsimumu}$ decays for muon identification, and ${\BKJPsiee}$ decays for electron identification. For the muon and hadron calibration samples, background contributions are subtracted using the \sPlot procedure~\cite{LHCb-DP-2018-001, Pivk:2004ty}, and PID efficiencies are computed in bins of the probe's transverse momentum and pseudorapidity. The \sPlot procedure is not applicable to the electron calibration sample due to correlations between the momentum of the probe electron and reconstructed \Bp mass, as the mass resolution is strongly affected by the electron energy loss before the calorimeter.
Instead, the fraction of candidates passing the electron PID criteria is estimated by fitting the invariant-mass distribution of the ${\BKJPsiee}$ data twice: once with no PID selection applied and again with the electron PID selection applied to the probe. This is repeated in bins of the probe's transverse momentum, pseudorapidity, and whether or not bremsstrahlung is recovered for the probe. 

The calibration of the PID efficiency implicitly assumes that the PID response of each final-state particle forming a \Bp candidate is independent of one another, with the exception of kinematic correlations. This assumption is tested using simulation samples and is found to have a negligible effect on the overall efficiency.

An initial set of corrections are derived to correct for the mismodelling of the \Bp production kinematics, impact parameter and vertex quality in the simulation. These corrections are obtained using the \BKJPsimumu channel and are applied to both muon and electron channels, after PID and tracking efficiencies have been corrected for. The corrections to the efficiencies discussed in the following are obtained on top of these initial corrections to the simulation.

The largest efficiency difference between muons and electrons originates from the trigger, where the $\et$ requirement for electrons is significantly more restrictive than the muon \pt requirement. 
The normalisation channels are employed to compare the trigger response between data and simulation using the tag-and-probe method. For example, using \BKJPsiee decays, the tag kaon is required to pass a set of trigger requirements. The efficiency of the electron trigger is then determined by the fraction of candidates in which at least one of the two probe electrons passes the electron trigger requirements. This efficiency is compared between data and simulation as a function of the maximum $\et$ of the two electrons, which is expected to be the kinematic variable on which the electron trigger efficiency most strongly depends. The efficiency for the muon trigger is determined similarly, but as a function of the \pt of the muon. 

\begin{figure}[tb]
  \begin{center}
    \centering
    \includegraphics[page=1]{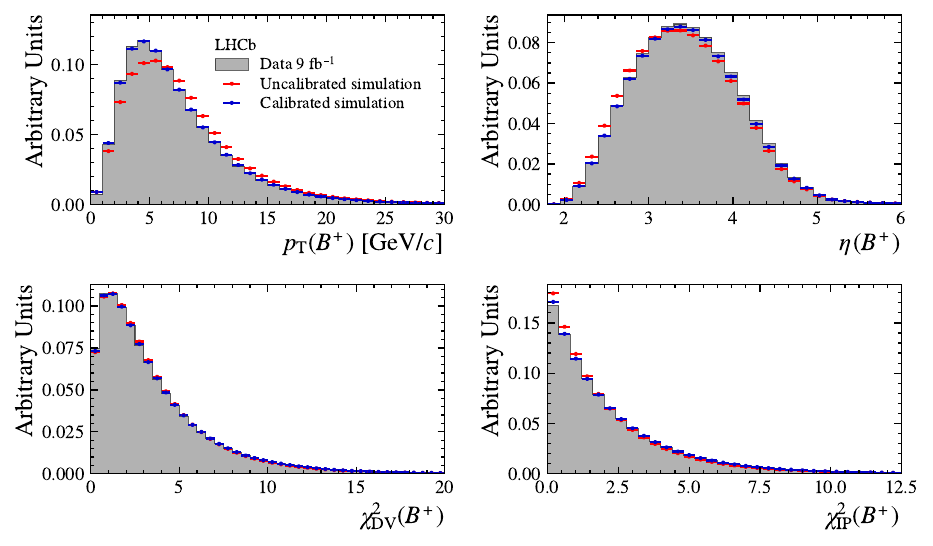}
    \vspace*{-0.5cm}
  \end{center}
  \caption{
    \small 
    Distributions of reconstructed \Bu meson properties for (grey) \BKJPsimumu data compared to simulation (red) before and (blue) after corrections are applied.} 
  \label{fig:data_sim_comparison}
\end{figure}

An additional correction is obtained to reduce any residual mismodelling of the $\Bp$ production kinematic and vertex-reconstruction quality. A gradient boosted reweighting procedure~\cite{Rogozhnikov:2016bdp} is applied as a function of $\pt(\Bp)$, $\eta(\Bp)$, $\chiSqBVtx$ and $\ipChiSq(\Bp)$ variables. The quantity $\chiSqBVtx$ represents the fit quality of the \Bp decay vertex, whereas $\ipChiSq(\Bp)$ is defined as the difference in the vertex-fit \chisq of the PV when reconstructed with and without the \Bp candidate. These corrections are obtained using the \BKJPsimumu channel and are assumed to be independent of the lepton species. 
 A comparison of the \BKJPsimumu data and simulation, both before and after the corrections, is presented in \figref{fig:data_sim_comparison}.

The quantity \qsqTrack, which is used to select both the electron normalisation and the rare channel, is known to be incorrectly modelled in the simulation. For the normalisation modes' efficiencies, a correction is therefore derived using \BKJPsiee simulation and data samples. The invariant-mass distribution of the electron modes depends significantly on the outcome of the bremsstrahlung recovery algorithm. Therefore, the samples are split into the corresponding $0\gamma$, $1\gamma$, and $2\gamma$ categories, depending on whether photons are added to none, one, or both electron candidates. For each bremsstrahlung category the \qsqTrack distribution in simulation is fitted with a combination of Crystal Ball functions~\cite{Skwarnicki:1986xj}. The resulting models are then convolved with a Gaussian function, and the corresponding \BKJPsiee data is fitted, with peak position and standard deviation determined from this fit. The extracted Gaussian function is then used to smear the \qsqTrack variable in the simulation, improving the agreement in resolution between the simulation and data.

The efficiencies computed using simulation are validated by measuring the branching fraction ratios of the resonant control modes, which have been shown to be consistent with LU~\cite{PDG2024}. The first of these is the single-ratio test, defined as:
\begin{equation}\label{rjpsi_single_ratio}
    \begin{aligned}
        \rJPsi &\equiv \frac{\BR(\BKJPsimumu)}{\BR(\BKJPsiee)}\\
        &= \frac{N(\BKJPsimumu)}{\varepsilon(\BKJPsimumu)}\cdot\frac{\varepsilon(\BKJPsiee)}{N(\BKJPsiee)}.
    \end{aligned}
\end{equation}
In this expression, $N(X)$ is the measured yield for decay $X$, while $\varepsilon(X)$ denotes the corresponding selection efficiency, which incorporates the effects of all the requirements outlined in \secref{sec:selection}. The large branching fraction for \BKJPsill decays means that \rJPsi is expected to have a small statistical uncertainty. Furthermore, since \rJPsi is a single-ratio, obtaining a result consistent with unity requires accurate control of the muon and electron efficiencies.

Additionally, the double-ratio between the \psitwos and \jpsi control modes,
\begin{equation}\label{RPsi2S_double_ratio}
    \begin{aligned}
        \RPsiS &\equiv \frac{\BR(\BKpsiSmumu)}{\BR(\BKpsiSee)} \cdot \frac{\BR(\BKJPsiee)}{\BR(\BKJPsimumu)} \\
        &= \frac{N(\BKpsiSmumu)}{\varepsilon(\BKpsiSmumu)}\cdot\frac{\varepsilon(\BKpsiSee)}{N(\BKpsiSee)} \cdot \frac{1}{\rJPsi},
    \end{aligned}
\end{equation}
is measured. Since \RPsiS is a double-ratio and both the \psitwos and \jpsi control mode datasets are large, this measurement is statistically precise with a small efficiency-related systematic uncertainty. The sources of systematic uncertainty associated with the selection efficiencies appearing in~\equref{rjpsi_single_ratio} and~\equref{RPsi2S_double_ratio} are presented in \secref{sec:systs}.  Resonant \BKpsiSmumu candidates are required to have $12.5<\qsq<14.2\gevgevcccc$, whereas \BKpsiSee candidates are selected with $10.5<\qsqTrack<14.2\gevgevcccc$. The latter requirement significantly reduces backgrounds from \BKJPsiee decays that would otherwise leak into the \BKpsiSee data if \qsq is used. Resonant \BKpsiSll candidates are selected with $5180<\mkllconstpsitwos<5600\mevcc$, where \mkllconstpsitwos is the reconstructed \Bp mass computed with the mass of the dilepton system constrained to the known \psitwos mass~\cite{PDG2024}.

\begin{figure}[tb]
  \begin{center}
    \centering
    \includegraphics[width=0.48\textwidth,keepaspectratio]{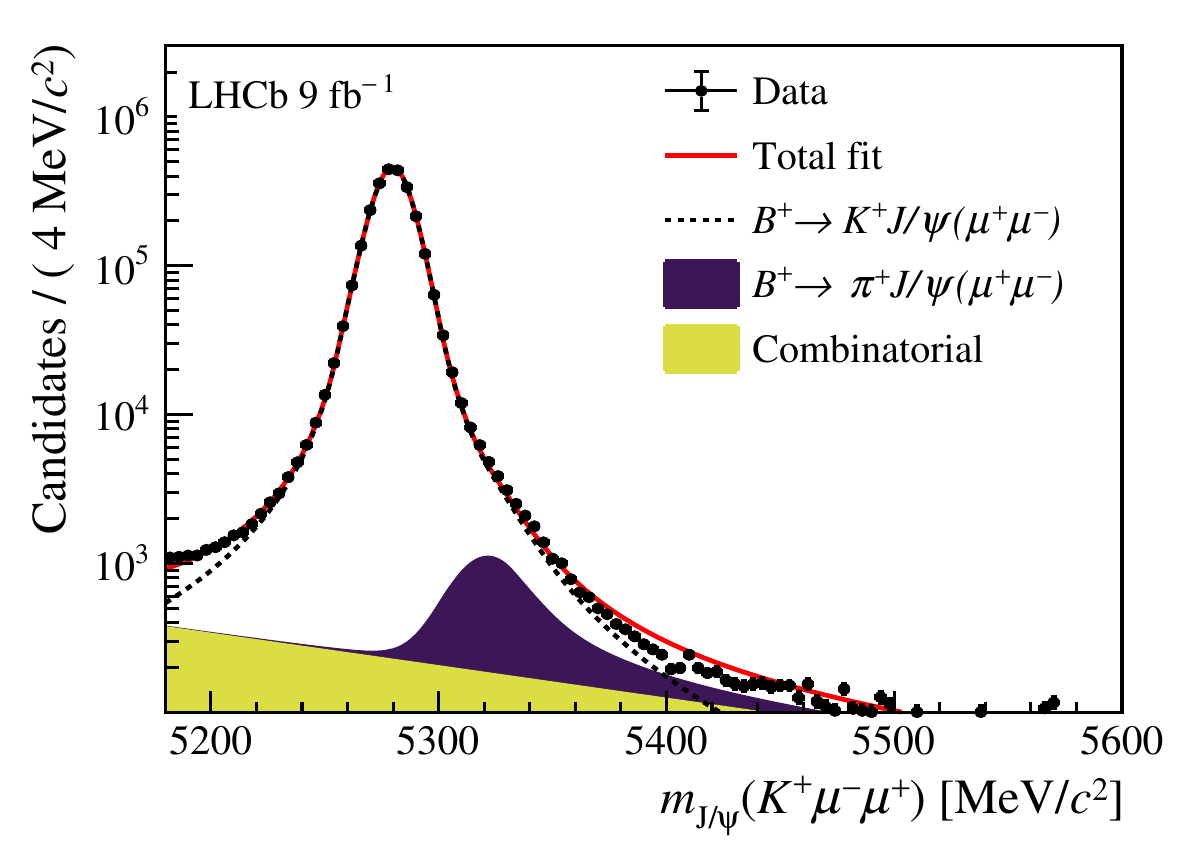}
    \includegraphics[width=0.48\textwidth,keepaspectratio]{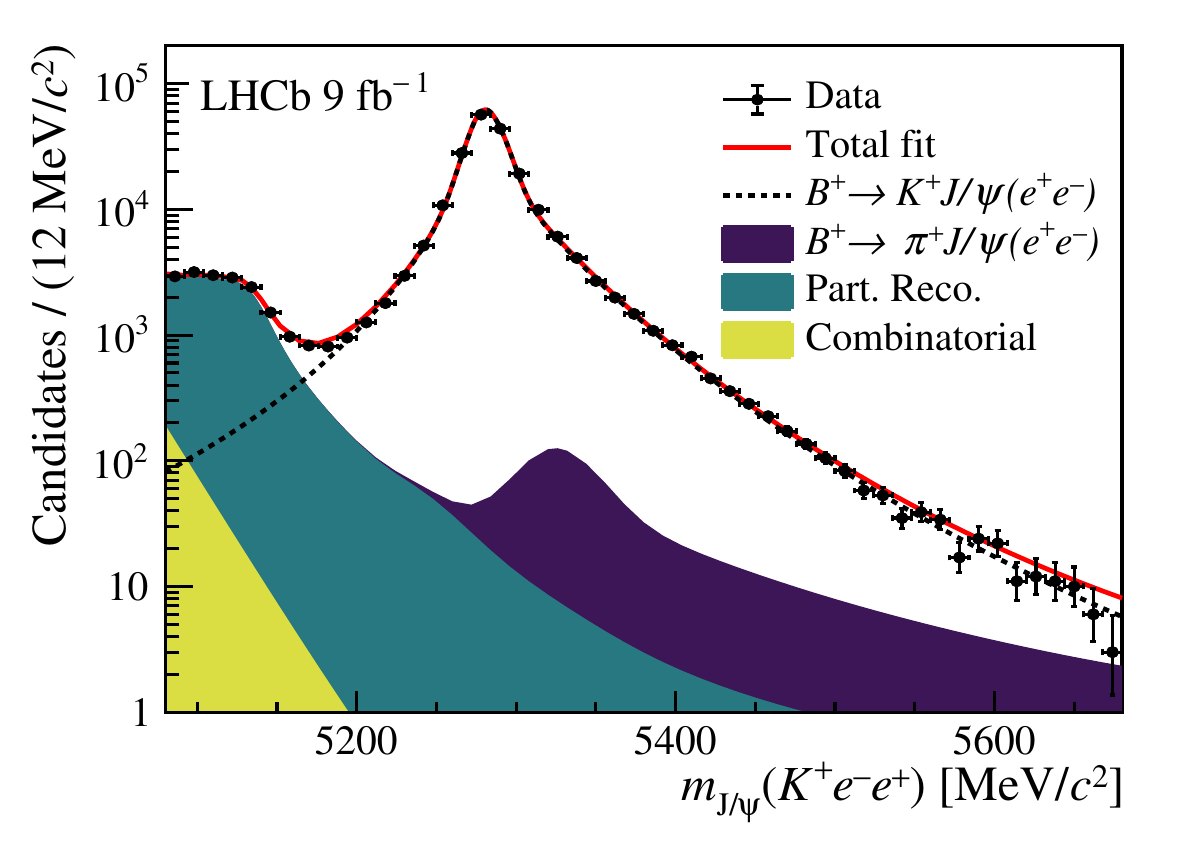}
    \includegraphics[width=0.48\textwidth,keepaspectratio]{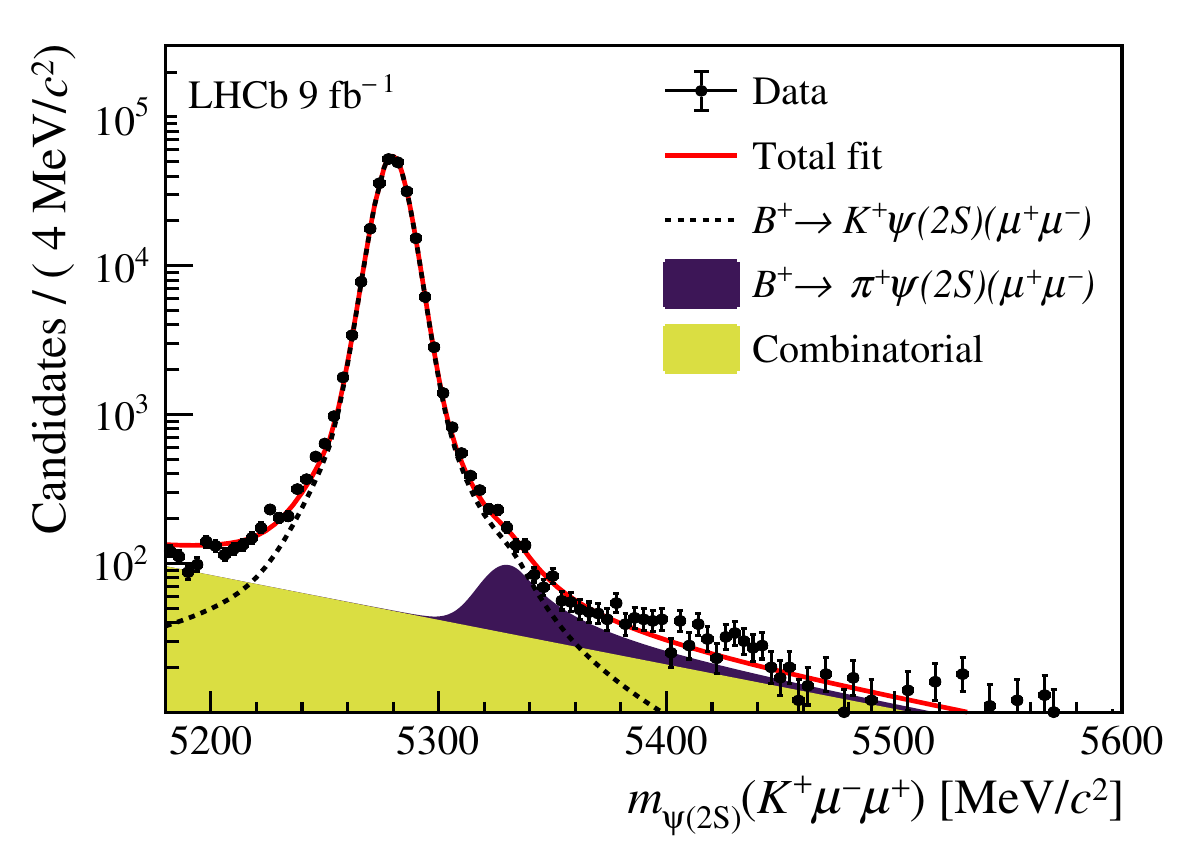}
    \includegraphics[width=0.48\textwidth,keepaspectratio]{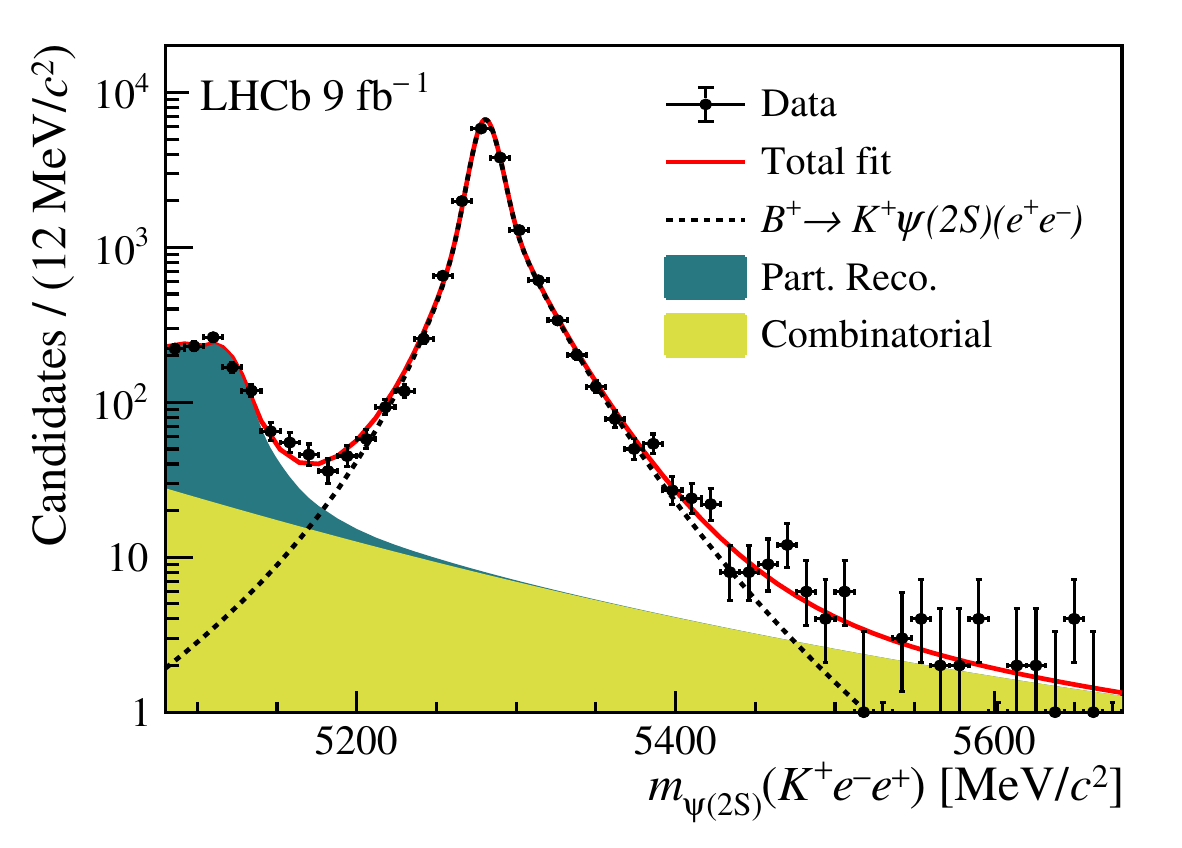}
    \vspace*{-0.5cm}
  \end{center}
  \caption{
    \small 
    Distributions of (top) \mkllconst of the \BKJPsill channel and (bottom) \mkllconstpsitwos of the \BKpsiSll channel for the (left) muon and (right) electron modes, and corresponding fit results.}
  \label{fig:roger}
\end{figure}

The yields of the control modes are determined using extended unbinned maximum-likelihood fits to the dilepton-constrained invariant-mass distribution. A breakdown of the selection efficiencies and yields for the control channels is reported in \tabref{tab:yieldsandeffs}. The fit results are shown in \figref{fig:roger}.  In the \mkmumuconst and \mkmumuconstpsitwos invariant-mass distributions, the small disagreement near $5400\mevcc$ has a negligible impact on the signal-yield determination. Moreover, contributions of misidentified \BPipsiSee decays where the final-state pion is misidentified as a kaon, are found to be negligible in \BKpsiSee decays. The resulting control mode yields are combined with efficiencies calculated using the corrected simulation and give

\begin{equation*}
    \begin{aligned}
        \rJPsi & = 0.997\pm0.003\pm0.055, \\
        \RPsiS & = 1.002\pm0.009\pm0.004,
    \end{aligned}
\end{equation*}
where the first uncertainty is statistical and the second is systematic (see Sec.~\ref{sec:systs}). Both cross-checks are consistent with the SM expectation of unity. 

\begin{figure}[tb]
  \begin{center}
    \centering
    \includegraphics[width=0.48\textwidth,keepaspectratio]{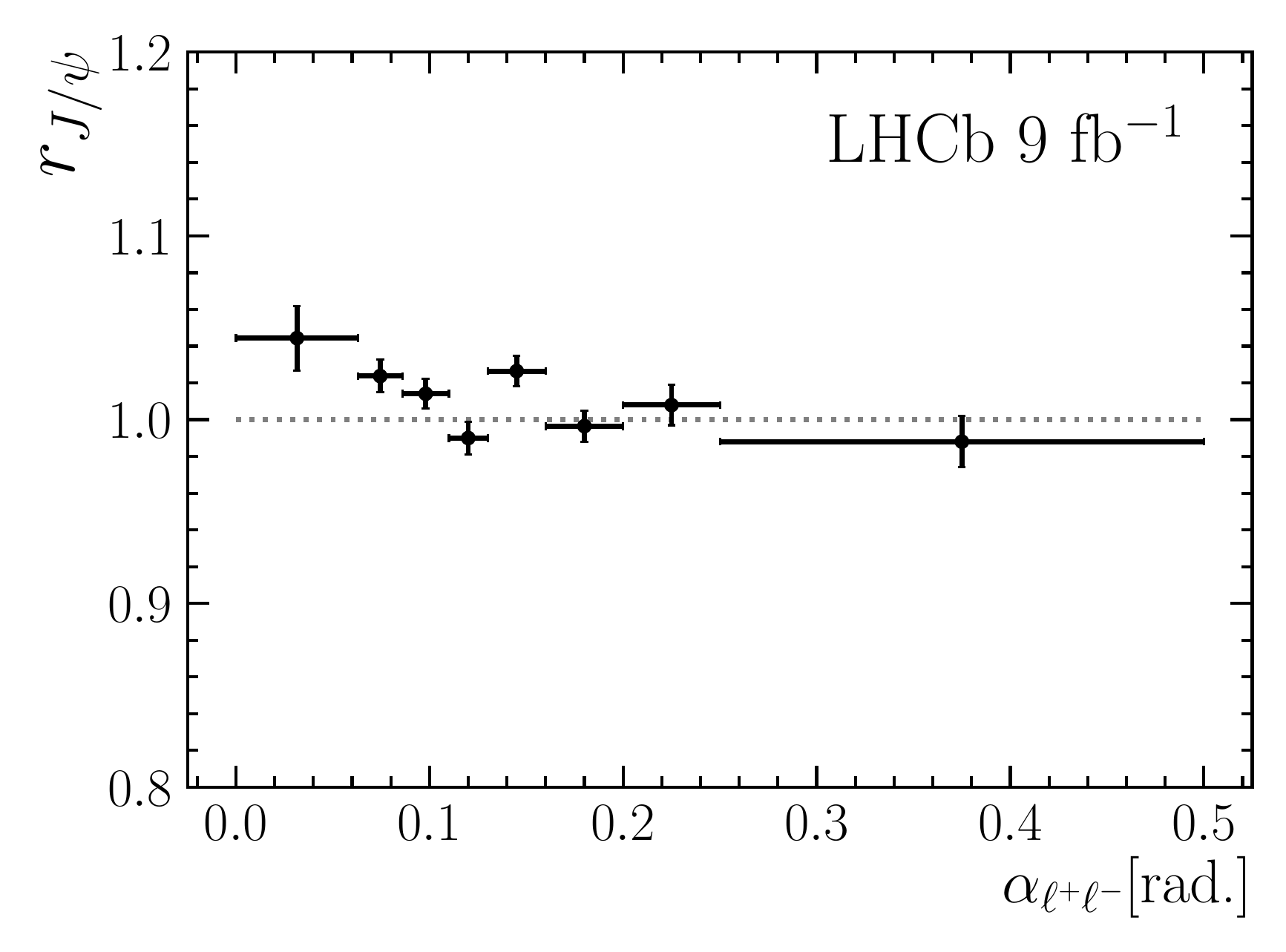}
    \includegraphics[width=0.48\textwidth,keepaspectratio]{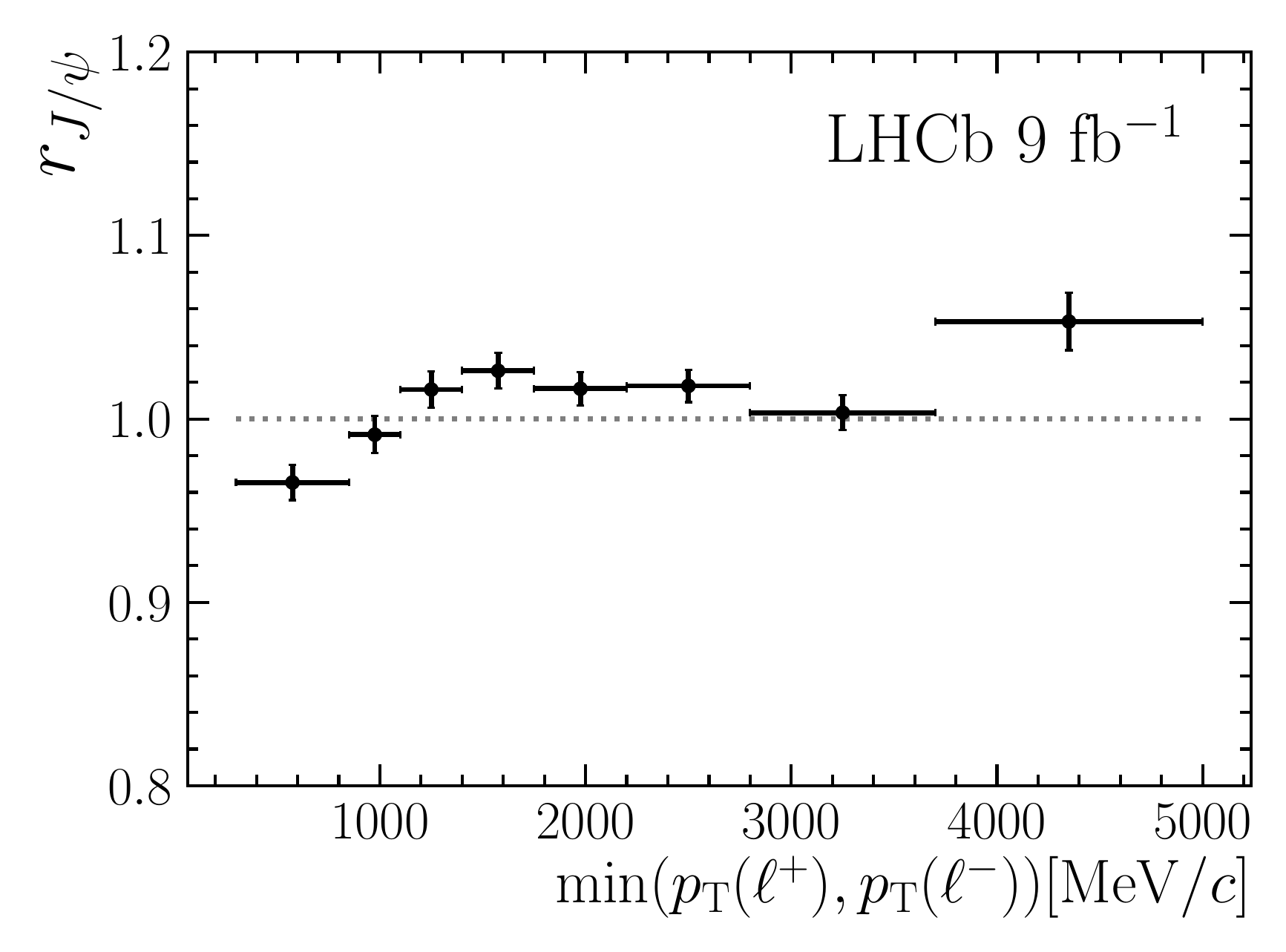}
    \vspace*{-0.5cm}
  \end{center}
  \caption{
    \small 
    Value of \rJPsi as a function of (left) the opening angle between the leptons $\alpha_{\ellp\ellm}$ calculated in the laboratory frame, and (right) the minimum \pt of the leptons. The error bars shown are statistical only.}
  \label{fig:diff_rJpsi}
\end{figure}

The distributions of variables, such as the opening angle and minimum transverse momentum of the final-state leptons, are expected to depend strongly on \qsq. Incorrectly modelling these distributions can result in biases in the efficiencies that do not cancel well in the double-ratio. Therefore, the single-ratio \rJPsi is measured in intervals of these variables (see~\figref{fig:diff_rJpsi}); the absence of significant variations in the observed distributions demonstrates sufficient control of the efficiencies for an unbiased measurement of \RK. The uncertainties in \figref{fig:diff_rJpsi} are statistical only, while efficiency-related systematic uncertainties (see \secref{sec:systs}) are sufficient to explain the residual deviations from uniformity. 

\begin{table}[tb]
\centering
\caption{\small Yields of the resonant channels obtained from the fits to the data and the corresponding efficiencies determined using calibrated simulation samples. The quoted uncertainties account for statistical effects only.} \label{tab:yieldsandeffs}
\begin{tabular}{lcc}
\toprule
Decay mode  &   Yield  & Efficiency  (\textperthousand)         \\
\midrule
\BKJPsiee    &   $\phantom{0\,}203\,000 \pm \phantom{0\,}450$   &  $\phantom{0}1.388  \pm 0.002$ \\
\BKJPsimumu  &   $2\,280\,000 \pm 1\,500$                       &  $15.680 \pm 0.015$ \\
\BKpsiSee    &   $\phantom{00\,}15\,700 \pm \phantom{0\,}130$   &  $\phantom{0}1.325  \pm 0.003$ \\
\BKpsiSmumu  &   $\phantom{0\,}201\,300 \pm \phantom{0\,}450$   &  $17.051 \pm 0.018$ \\
\bottomrule
\end{tabular}
\end{table}

\section[\RK measurement method]{\boldmath \RK measurement method}\label{sec:fits}

Similar to the formulation of \RPsiS in \equref{RPsi2S_double_ratio}, \RK can be measured using the double-ratio
\begin{equation} \label{eq:RK}
\RK = \dfrac{N(\BKmumu)}{\varepsilon(\BKmumu)} \cdot \dfrac{\varepsilon(\BKee)}{N(\BKee)} \cdot \dfrac{1}{\rJPsi}. 
\end{equation}
The terms $\varepsilon(\BKee)$ and $\varepsilon(\BKmumu)$ represent the selection efficiencies of rare \BKee and \BKmumu decays, respectively. These are model-dependent quantities as they are computed by integrating over a model of the \qsqTrue distribution of the decay, where ${\qsqTrue \equiv (p_{B^{+}} - p_{K^{+}})^{2}}$ is the squared momentum transferred to the dilepton pair. Here and in the following, the subscript $\it{true}$ is used to indicate quantities prior to any final state radiation and detector resolution effects. This model dependence is problematic since multiple broad open-charm resonances exist in the high-\qsq region, as well as interference effects with the upper tail of the nearby large $\psi(2S)$ resonance, both of which are not precisely known. 

\begin{figure}[!tb]
  \begin{center}
    \centering
    \includegraphics[page=1]{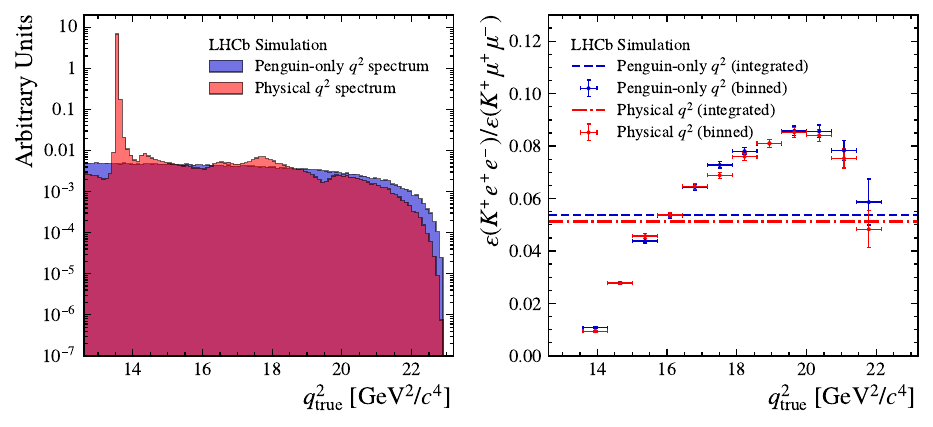}
    \vspace*{-0.5cm}
  \end{center}
  \caption{
    \small (left) Comparison of the $\qsqTrue$ distribution and (right) the ratio of efficiencies between muonic and electronic rare mode decay channels for two different $\qsqTrue$ models. The ratio of efficiencies is shown for both a binned and integrated calculation (dashed lines). The \qsqTrue model is varied from a simplistic description that considers only rare mode decays (blue) to a more sophisticated model that includes resonant processes and interference effects (red).}
  \label{fig:model_independent}
\end{figure}

To demonstrate the effect of a large mismodelling on the efficiencies, an example is considered using two significantly different \qsqTrue spectra. The baseline distribution of \qsqTrue is obtained from a measurement using \BKmumu decays~\cite{LHCb-PAPER-2016-045}, and is shown as the red histogram on the left-hand side of \figref{fig:model_independent}. In contrast, the blue histogram represents a simplified \qsqTrue model~\cite{Ball:2004ye}, which accounts only for penguin contributions to the decay amplitude. The ratio of rare-mode integrated selection efficiencies, $\varepsilon(\BKee)/\varepsilon(\BKmumu)$, is calculated using each model and illustrated by the horizontal lines on the right-hand side of \figref{fig:model_independent}. In this extreme example, the ratio shifts by approximately 5.5\%, depending on the \qsqTrue model used, which would directly translate to a bias in \RK if measured using \equref{eq:RK}. This motivates using a less model-dependent method for determining \RK. The method described in this paper reduces the uncertainty associated with altering the \qsqTrue model in this extreme example to at most 0.5\%.

To reduce the \qsqTrue model dependence, one can exploit cancellations in the ratio of efficiencies in \equref{eq:RK}. In the scenario that both the \qsqTrue spectra and the \qsqTrue dependence of the selection are identical between the muon and electron modes, any model dependence cancels in the ratio of efficiencies. In the case of LU, as in the SM, the \qsqTrue spectra of each mode are identical. However, the \qsqTrue dependence of the selection efficiencies of each mode is dramatically different, primarily due to the \qsqTrack requirement used to select \BKee decays. 
This can be circumvented by weighting the muon data as a function of \qsq, such that the \qsqTrue dependence of the muon efficiency aligns with that of the electron mode.
 The weights applied to the rare muon data to achieve this alignment, $w_{\varepsilon}(\qsqTrue)$, are defined as the ratio of the muon and electron efficiencies as a function of \qsqTrue as shown on the right-hand side of \figref{fig:model_independent}. For each of the previously considered \qsqTrue models, the $w_{\varepsilon}(\qsqTrue)$ distribution changes only by a small amount when the \qsqTrue model is varied significantly, with the residual variation attributed to the limited size of the simulation samples and the finite binning. Furthermore, there is residual model dependence arising from the assumption that, for \BKmumu decays, \qsq and \qsqTrue are equivalent. The impact of this assumption on \RK is minimal, and a more detailed discussion is presented in~\secref{sec:systs}.

Before applying the weights $w_{\varepsilon}(\qsqTrue)$ to the rare muon data, backgrounds are subtracted using the \sPlot technique~\cite{LHCb-DP-2018-001}. The \sPlot method is performed by fitting the \mkmumu distribution of the rare muon data, as described in \secref{sec:muon_fit}. Each candidate in the rare muon data, indexed by $i$, is then represented by an \emph{sWeight}, $s\mathcal{W}_i$. Using the \emph{sWeights}, \RK is determined as
\begin{equation} \label{eq:RK_model_independent}
\RK = \underbrace{\dfrac{1}{\rJPsi} \cdot \sum_{i} s{\mathcal{W}_i} \cdot  w_{\varepsilon}^{i}(\qsqTrue)}_{C}\cdot \dfrac{1}{N(\BKee)},
\end{equation}
where the sum over $i$ corresponds to the sum over fully selected candidates in the rare muon data. Importantly, unlike the fit strategy described in Refs.~\cite{LHCb-PAPER-2022-045,LHCb-PAPER-2022-046}, \RK is not the result of a simultaneous fit to rare electron and rare muon data. Instead, \RK is extracted in the following manner: all efficiency terms, resonant channel yields, and information from the rare muon data are combined into the factor $C$, which corresponds to the expected yield of rare electron decays when $\RK = 1$. The yield of \BKee decays, denoted $N(\BKee)$, is determined by performing an extended unbinned maximum-likelihood fit to the \mkee distribution of the rare electron data, detailed in \secref{sec:electron_fit}. The likelihood function is parametrised in terms of $C$ and \RK, allowing \RK to be determined directly from the fit. A Gaussian term is included in the likelihood function to constrain $C$ to its estimated value, $C = 198.8 \pm 3.7 \stat\pm 2.2 \syst$, as determined from data and simulation. This ensures that uncertainties related to efficiencies, resonant channel yields, and rare muon data are included in the uncertainty on \RK.

\subsection{Muon rare mode fit}\label{sec:muon_fit}
An extended unbinned maximum-likelihood fit is performed to the \mkmumu distribution of the rare muon data as part of the \sPlot procedure, used to derive \emph{sWeights}. The fit model consists of three components: combinatorial background, \BKpsiSmumu decays that leak up into the high-\qsq region due to resolution effects, and rare \BKmumu signal decays. 

The signal peak model is derived in the following manner. For each data-taking year separately, the $m(\Kmumu)$ distribution in the \BKmumu simulation is fitted with a sum of two Crystal Ball functions and a Gaussian distribution with a shared peak position but independent width parameters. The peak position and the global width of each distribution are corrected for data-simulation differences in the mass resolution, using a correction derived from the \BKJPsimumu resonant channel. The parameters from the fits to simulation are then fixed given their negligible uncertainties, and the distributions for each data-taking year are combined in a weighted sum. 

The shape of the \BKpsiSmumu background is modelled using the sum of a Crystal Ball function and a Gaussian distribution. The parameters of the model are determined from fitting the \BKpsiSmumu simulation samples and then fixed in the fit to the data. The yield of the \BKpsiSmumu component is constrained using a Gaussian penalty function to an estimate obtained by scaling the fitted \BKpsiSmumu yield from the resonant mode fits (see \figref{fig:roger}) by the ratio of the rare and resonant \qsq selection efficiencies obtained from \BKpsiSmumu simulation.

\begin{figure}[tb]
  \begin{center}
    \centering
    \includegraphics[width=0.65\textwidth,keepaspectratio]{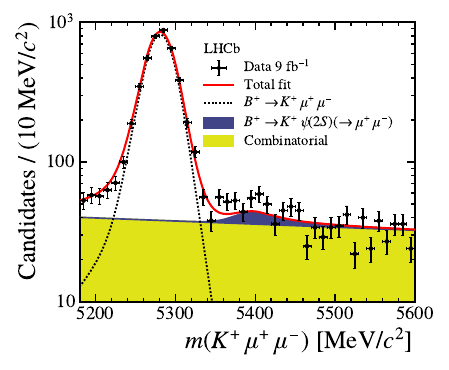}

    \vspace*{-0.5cm}
  \end{center}
  \caption{
    \small Distribution of \mkmumu for the rare muon data compared to the result of the fit described in the text. The \BKpsiSmumu background peaks at a shifted position with respect to the rare mode due to the \qsq requirement of being larger than the \psitwos squared mass.}
  \label{fig:fit_result_Kmumu}
\end{figure}

An exponential function is used to model the combinatorial background, with its normalisation and slope parameter allowed to float freely in the fit. The fit to the rare muon mode data is shown in~\figref{fig:fit_result_Kmumu}. The fitted yield of \BKmumu decays is $N(\BKmumu) = 4\,008 \pm 71$.

\subsection{Electron rare mode fit}\label{sec:electron_fit}
The yield of \BKee decays, $N(\BKee)$, is extracted by performing an extended unbinned maximum-likelihood fit to the $m(\Kee)$ distribution of the rare electron data. The modelling of each component contributing to the fit is discussed below.

\subsubsection{Signal peak}
The invariant-mass distribution of the electron rare mode has a significant dependence on the bremsstrahlung category, therefore, the samples are split according to bremsstrahlung category and each resulting distribution is described by a different shape. Category $0\gamma$ is modelled by a sum of two Crystal Ball functions, with opposite tails with respect to the peak. Two Gaussian distributions are also included, one to improve the modelling of the low-mass tail and the other to improve the modelling of the core of the distribution. Similarly, category $1\gamma$ is modelled by the sum of two Crystal Ball functions and a Gaussian distribution that improves the modelling of the distribution's core. Finally, category $2\gamma$ is modelled with the sum of two Crystal Ball functions with tails on both sides of the peak. The shape parameters, initially obtained by fitting the simulated \mkee distributions, are then corrected for data-simulation differences in resolution for each bremsstrahlung category. This procedure is analogous to the one adopted for the electron control modes (as described in \secref{sec:eff}), however, due to the high-\qsq requirement on the electron rare mode, in this case the corrections are derived using simulation and data samples of \BKpsiSee decays. The three smeared distributions are then summed with a fraction obtained from \BKee simulated events. Due to the use of \qsqTrack in the data selection, only a small fraction of decays are expected in 2$\gamma$ (5.4\%) compared to 0$\gamma$ (54.7\%) and 1$\gamma$ (39.9\%) categories. It is checked that the simulation accurately reproduces these fractions by comparing their values in data and simulated control modes. In the fit to the rare electron data, all the parameters associated with the description of the signal invariant-mass shape are fixed.

\subsubsection{Partially reconstructed background}
The most significant sources of partially reconstructed backgrounds are from $\Bu$ and $\Bd$ meson decays. The largest contributions are ${\decay{ \Bz }{ \Kstar(892)^0(\to \Kp\pim)\epem}}$ and ${\decay{ \Bu } { \Kstar(892)^+(\to \Kp\piz)\epem}}$ decays, followed by much smaller contributions that involve other intermediate states. The simulation samples for  $\decay{ \Bz }{ \Kstar(892)^{0}(\to \Kp\pim)\epem}$ and $\decay{ B^+ }{ \Kstar(892)^{+}(\to \Kp\pi^0)\epem}$ modes include $K\pi$ \mbox{$P$-wave} resonant decays only. Therefore, the $m(K\pi)$ spectrum for these samples is weighted to account for $K\pi$ $S$-wave contributions using the lineshape and the measured normalisation of Ref.~\cite{LHCb-PAPER-2016-012}. The simulation samples for each decay mode are combined into a single inclusive sample. Each process contributing to the inclusive sample is weighted to take into account its branching fraction and selection efficiency. The resulting inclusive sample is used to build a nonparametric Kernel Density Estimation (KDE), which is used as the model for the partially reconstructed background component in the rare electron mode fit. 
The shape is corrected for data-simulation differences in the $m(\Kee)$ resolution in a similar manner as the signal decay. The yield of the partially reconstructed background component is allowed to float freely in the fit to the data. An additional partially reconstructed background, due to $\decay{ \Bs} {\phi(\to \Kp\Km)\epem}$ decays, is expected to make a small contribution of only five candidates~\cite{PDG2024}. This contribution is modelled similarly to the previous component; however, its yield is constrained in the fit. 

\subsubsection{Combinatorial background}
The selection of candidates with large dilepton invariant mass enforces a lower bound on the possible value of $\mkee_{\rm{true}}$, which significantly impacts the combinatorial background that would otherwise follow an exponential distribution. The effect of this restriction on the $\mkee_{\rm{true}}$ distribution of the combinatorial background is modelled using the three-body phase-space function

\begin{equation}\label{eq:phsp}
   \deriv \phi_{3}(x, m_{K}, q_{\rm{true}} ) = \kappa(x, m_{K}, q_{\rm{true}}) \,  \deriv x \, \deriv q_{\rm{true}} ,
\end{equation}
\noindent where $\kappa(x, m_{K}, q_{\rm{true}})$ is the triangle function

\begin{equation}\label{eq:kallen}
   \kappa(x, m_{K}, q_{\rm{true}}) = \frac{\sqrt{x^{2}-(m_{K}+q_{\rm{true}})^{2}}}{x} \frac{\sqrt{x^{2}-(m_{K}-q_{\rm{true}})^{2}}}{x} ,
\end{equation}

\noindent which depends on the kaon mass, $m_K$, the invariant mass squared of the dilepton pair, $q_{\rm{true}}$, and the three-body invariant mass $x\equiv\mkee _{\rm{true}}$. The two-body term $\deriv \phi_{2}(q_{\rm{true}}; p_{\ell^{+}}, p_{\ell^{-}})$ is neglected, since at high-dilepton invariant mass it is very close to one.  The reconstructed mass distribution describing the combinatorial component at high-dilepton invariant mass is obtained by making the ansatz

\begin{equation}\label{eq:combipdf}
\frac{\deriv \Gamma}{\deriv x}(x, q^{\textrm{min}}_{\rm{true}} , q^{\textrm{max}}_{\rm{true}}, \lambda) = \frac{1}{4 x } e^{-\lambda x}\int_{q^{\textrm{min}}_{\rm{true}}}^{q^{\textrm{max}}_{\rm{true}}}   \kappa(x,m_{K},q_{\rm{true}} ) \, \deriv q_{\rm{true}}
\end{equation}

\noindent where $q^{\textrm{min}}_{\rm{true}}$ and $q^{\textrm{max}}_{\rm{true}}$ are the kinematic boundaries of the measurement, and $\lambda$ is the exponential slope. This model is hereafter referred to as the phase-space model. 

\begin{figure}[tb]
  \begin{center}
    \centering
    \includegraphics[page=1]{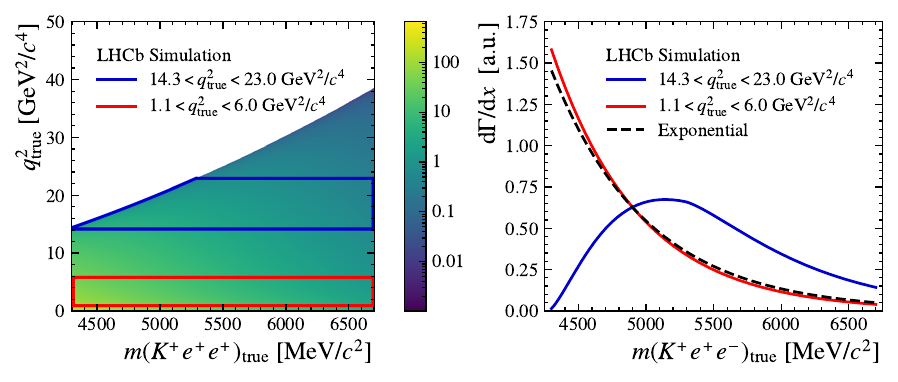}
    \vspace*{-0.5cm}
  \end{center}
  \caption{ 
    \small Visualisation of the phase space (left) in a two-dimensional representation of the integrand of \equref{eq:combipdf} parametrised in \qsqTrue and $\mkee _{\rm{true}}$ and (right) the phase-space model for two highlighted \qsqTrue regions.}
  \label{fig:phase_space}
\end{figure}

The integrand of \equref{eq:combipdf} parametrised in terms of \qsqTrue and $\mkee _{\rm{true}}$ is illustrated on the left-hand side of \figref{fig:phase_space}. Two different \qsqTrue regions are highlighted: one corresponding to the region $1.1 < \qsqTrue < 6.0 \gevgevcccc$, where the combinatorial background is expected to follow an exponential distribution, and the other corresponding to the high-\qsqTrue region probed by this analysis. The phase-space model for each \qsqTrue region is illustrated on the right-hand side of \figref{fig:phase_space}. The phase-space model for the region $1.1 < \qsqTrue < 6.0 \gevgevcccc$ is compatible with an exponential distribution with the same value of the slope parameter $\lambda$. Conversely, the phase-space model for the high-\qsqTrue region has the expected characteristic of decreasing in size as $\mkee_{\rm{true}}$ decreases towards 4300\mevcc.

The advantage of using a physics-motivated model to describe the combinatorial background lies in its ability to reduce the number of free parameters required to parametrise the shape. By fixing the parameters $q^{\textrm{min}}_{\rm{true}}$ and $q^{\textrm{max}}_{\rm{true}}$ to their expected values and only allowing $\lambda$ to freely float, the uncertainty on the extracted \BKee yield is reduced, which translates to a reduced \RK uncertainty.

The $\mkee_{\rm{true}}$ distribution of combinatorial candidates in data is impacted by detector effects, which are not accounted for in the derivation of the phase-space model. To account for this, the phase-space model is corrected using data control samples.

The first control sample used is the mixed sample, which is obtained by combining a \Kp, \ep and \en candidate from three randomly selected events from the data. For each new \Kee candidate, kinematic quantities such as $m(\Kee)$, $\qsqTrack$ and $m(\Kp\en)$ are computed. The mixed sample is used to derive corrections to the phase-space model to account for the sculpting from the charm-cascade kinematic veto $m(\Kp\en)>1885 \mevcc$, and the assumption in the derivation of the phase-space model that a cut is made on \qsqTrue, whereas in reality, the selection is based on $\qsqTrack$. 
Both effects act to exaggerate the sculpting, removing more candidates with a low value of \mkee.

\begin{figure}[tb]
  \begin{center}
    \centering
    \includegraphics[page=1]{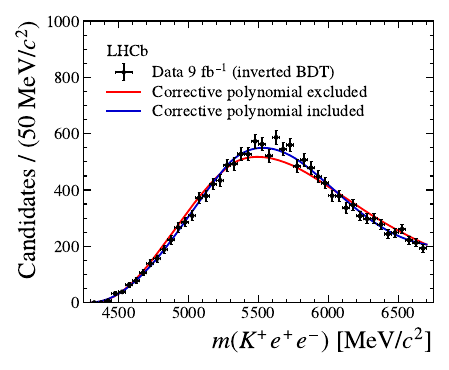}
    \includegraphics[page=1]{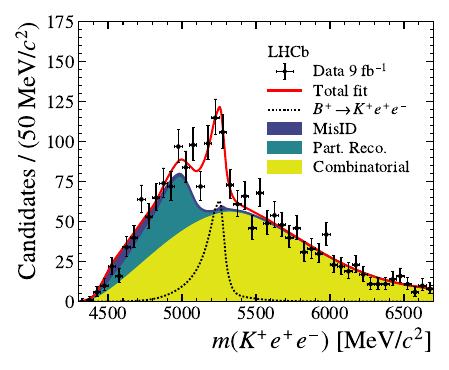}
    \vspace*{-0.5cm}
  \end{center}
  \caption{
    \small
    Fit projections comparing the combinatorial model to the rare electron mode data with (left) an inverted and (right) a loose cut on the combinatorial BDT output.}
  \label{fig:combinatorial_control}
\end{figure}

The model is then compared to the rare electron mode data sample with an inverted selection requirement on the combinatorial BDT response. The inverted BDT requirement ensures that the sample is enriched in combinatorial background and that signal-like backgrounds are suppressed to negligible levels. The combinatorial model is fit to the inverted BDT sample, with only the $\lambda$ parameter allowed to float. The fit result is illustrated by the red curve on the left-hand side of \figref{fig:combinatorial_control}. The minor discrepancies between the inverted BDT data and the fitted model are a result of mismodelling effect of the selection requirements on the ansatz of~\equref{eq:combipdf}. To account for this, the combinatorial model is multiplied by a polynomial function that reduces these dependencies, demonstrated by the blue curve. 

Finally, the BDT requirements have the effect of removing more candidates at higher values of $m(\Kee)$, and shifting the peak of the combinatorial distribution towards lower values of \mkee.
The \mkee distribution is corrected to account for this warping using a dedicated \BKemu control sample. This sample is expected to be dominated by combinatorial background, as \BKemu decays are forbidden in the SM~\cite{LHCb-PAPER-2019-022}. The selection applied to the \BKemu data is aligned as much as possible to that applied to the \BKee data, with the exception of the PID requirements applied to select a muon in the final state. Using this sample, the mass-dependent efficiency of the BDT selection is determined and incorporated into the total combinatorial model.

To validate the BDT correction derived from the \BKemu sample, a fit is performed to the electron rare mode data but with two modifications relative to the baseline fit used to determine \RK. First, all parameters associated with the shape of the combinatorial model are fixed to values obtained from the control samples. This differs from the baseline fit where the slope parameter $\lambda$ is allowed to vary freely. Second, the data is selected with a cut on the output of the combinatorial BDT that is looser than the one used in the baseline determination of \RK, resulting in a sample that is enriched in combinatorial background. However, the BDT requirement remains sufficiently tight that it still significantly impacts the \mkee distribution of the combinatorial background, the effect of which is re-evaluated using the \BKemu sample. The result of the fit is presented on the right-hand side of \figref{fig:combinatorial_control}. Despite the simplicity of the combinatorial model, good agreement is observed between the model and data, providing confidence that all relevant effects impacting the combinatorial background are well under control.

\subsubsection{Misidentified background}
Although PID selection requirements drastically suppress contributions from processes where a hadron is misidentified as an electron, it is necessary to estimate the size and shape of any residual misidentified background contributions, which fall into three main categories. The first two categories correspond to the fully reconstructed decays \BKpipinomisID and \BKKKnomisID. In the former decay, the two pions are misidentified as electrons, whereas in the latter, two kaons are misidentified as electrons. Accurately modelling these backgrounds is of particular importance since they both peak close to the \Bp mass. Hence, incorrectly modelling these backgrounds can significantly bias the extracted \BKee yield and, consequently, \RK. The invariant-mass parametrisations of these two components are modelled with KDEs obtained using dedicated simulation samples. 

\begin{figure}[tb]
  \begin{center}
    \centering
    \includegraphics[page=1]{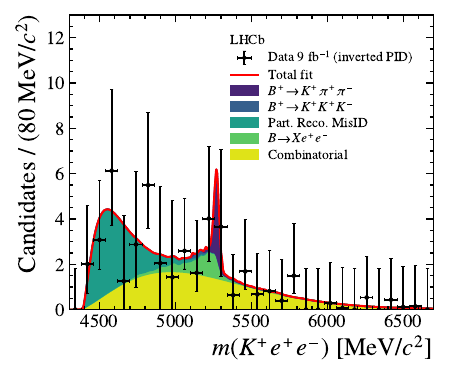}
    \includegraphics[page=1]{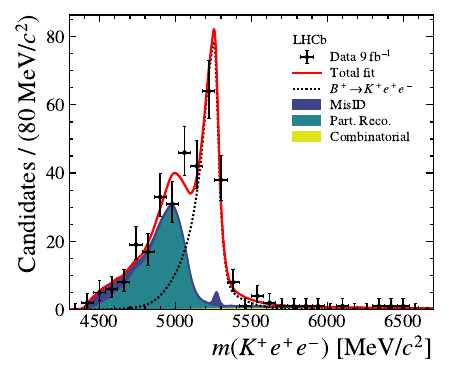}
    \vspace*{-0.5cm}
  \end{center}
  \caption{ 
    \small Invariant-mass distribution (left) of the inverted PID data control sample and (right) of the rare electron mode data, compared to the corresponding fits described in the text.}
  \label{fig:pass-fail}
\end{figure}

A third contribution arises from partially reconstructed misidentified backgrounds, which include any process with one or two hadrons misidentified as an electron and a particle that is not reconstructed as part of the $\Kp\ep\en$ final state. Due to the missing energy taken by the unreconstructed particle, candidates in this category are expected to form a broad peak at low values of \mkee compared to the \Bp mass. As there are many decays that contribute to this third category, most with imprecisely known branching fractions, it is impractical to model this background using simulated samples. Instead, an approach using data is utilised~\cite{LHCb-PAPER-2022-045}. An orthogonal control sample that is enriched in misidentified background is constructed by inverting the electron PID requirements and leaving all other criteria unchanged. Weights derived from standard PID calibration samples are used to scale and reshape the inverted PID data. A fit to the weighted data with inverted PID requirements enables an estimate of the misidentified background to be determined whilst removing contributions from combinatorial candidates and signal decays that leak into the inverted PID region. In this fit, KDEs built using dedicated simulation samples represent the \BKpipinomisID and \BKKKnomisID backgrounds along with a phase-space distribution with \qsqmin and $\lambda$ parameters floating to describe the nonpeaking misidentified background. Systematic uncertainties relating to this approach are assessed in \secref{sec:systs}, including one accounting for the fit model choice. The fit result is shown on the left-hand side of \figref{fig:pass-fail}. 

In the fit to the rare electron mode data, shown on the right-hand side of \figref{fig:pass-fail}, the shape and normalisation parameters of the misidentified background component are fixed. The estimated yields of each misidentified background, obtained using the data-based approach, are presented in \tabref{tab:misID_yields}. These are compared to alternative estimates derived using selection efficiencies computed from simulated samples reweighted to the measured Dalitz distribution and the measured branching fractions of \BKpipinomisID and \BKKKnomisID decays~\cite{PDG2024}. The uncertainty on the simulation-based estimates is primarily dominated by the precision of the \BKpipinomisID and \BKKKnomisID branching fractions and does not account for uncertainties related to the Dalitz distribution, which are expected to be significant. Despite this, the estimates based on simulation and data are consistent, providing confidence in the modelling of the misidentified background. 

\begin{table}[tb]
    \centering
    \caption{\small Estimated yield of the misidentified backgrounds computed using the baseline method based on data, and alternatively computed using selection efficiencies derived from simulation samples and branching fractions from Ref.~\cite{PDG2024}.}
    \label{tab:misID_yields}
    \begin{tabular}{lcc}
        \toprule
         & Data-based &  Simulation \\
        Background & \multicolumn{2}{c} {Yield}\\
        \midrule
        \BKpipinomisID & $\phantom{0}2.2 \pm 0.9$ & $1.24 \pm 0.09$  \\
        \BKKKnomisID   & $\phantom{0}0.2 \pm 0.4$ & $0.90 \pm 0.04$  \\
        Part. reco. misID  & $\phantom{.0}14 \pm 4\phantom{.0}$ & \text{--}  \\
        \bottomrule
    \end{tabular}
\end{table}

An additional cross-check is performed by measuring \RK under varying PID criteria, re-evaluating the misidentified background contribution at each step. A significant dependence of \RK on the PID requirements could indicate shortcomings in the modelling of the misidentified background contribution. To assess this, the observed PID scan is compared to pseudoexperiments generated under the null hypothesis, assuming no such dependence. No trend is observed in the pseudoexperiments, and a $p$-value of $0.80$ is obtained from the comparison. This result indicates good agreement between data and expectation, providing confidence in the robustness of the modelling of the misidentified background component.

\subsubsection{Fit validation}
The yield of \BKee decays, as determined from the fit to the rare electron data, is $N(\BKee) = 182^{+17}_{-16}$ where the uncertainty is statistical only. The stability of the fit is evaluated using pseudoexperiments generated from the baseline fit result shown by the red line on the right side of \figref{fig:pass-fail}. All pseudoexperiments consistently converge to a well-defined global minimum of the likelihood function, irrespective of the initial conditions of the fit.

\section{Systematic uncertainties}\label{sec:systs}
The most significant sources of systematic uncertainty stem from the modelling of the invariant-mass fit to rare electron data.
Additionally, there are systematic uncertainties relating to the correction chain applied to the simulated samples. Finally, there is a small source of systematic uncertainty arising from variations to the \qsqTrue model-independent approach for extracting \RK. All systematic uncertainties are expressed as percentages relative to the central fitted value of \RK. A summary of the systematic uncertainties affecting \RK is presented in \tabref{systematics:table}.

\begin{table}[tb]
    \centering
    \caption{\small Summary of relative systematic uncertainties on \RK, as described in the text.}
    \label{systematics:table}
    \begin{center}
        \begin{tabular}{lc}
            \toprule
            Source                             &  $\sigma_{\RK} (\%)$\\
            \midrule
            Fit bias                               &    1.1 \\
            Signal model                           &    0.9 \\ 
            Partially reconstructed background     &    2.2 \\ 
            Combinatorial background               &    1.2 \\ 
            Misidentified background               &    2.0 \\ 
            Excluded backgrounds                   &    0.6 \\ 
            \qsqTrue model-independent method      &    0.6 \\ 
            Efficiency corrections                 &    0.9 \\ 
            \bottomrule
        \end{tabular}
    \end{center}
\end{table}

A large ensemble of pseudoexperiments generated from the baseline fit result are used to estimate the intrinsic bias on \RK, which is determined to be $1.1\%$. The bias is corrected in the final \RK result and is further included as a systematic uncertainty.  Additional systematic uncertainties related to the modelling of components contributing to the invariant-mass spectrum are evaluated by fitting each pseudoexperiment twice: once with the \emph{baseline} model and once with an \emph{alternative} model. The \RK residuals, $\Delta\RK = R_K^{\rm{alternative}} - R_K^{\rm{baseline}}$, are then calculated for the ensemble of pseudoexperiments. For each source of uncertainty, the largest of the mean or standard deviation of the distribution of residuals is used as the systematic uncertainty. In cases where there are multiple sources of systematic uncertainty associated with a single fit component, the figures in \tabref{systematics:table} represent the sum in quadrature of all sources.

Model choices related to the signal peak in the fit to the rare electron data result in a $0.9\%$ systematic uncertainty on \RK, which accounts for the following variations. Firstly the effect of varying the signal peak parametrisation is evaluated by instead using a KDE constructed using the \BKee simulation. Secondly, the uncertainty associated with the data-based correction to the \mkee resolution of the signal peak is evaluated by using a signal peak excluding this correction. Finally, uncertainty associated with the limited size of the simulation sample used to derive the signal peak is evaluated by bootstrapping~\cite{Efron:1979bxm} the simulation samples for each pseudoexperiment and building a fresh KDE for each fit.

Systematic uncertainties related to the modelling of the partially reconstructed background component have a combined size of $2.2\%$. Similarly to the signal peak, the systematic uncertainty associated with the partially reconstructed background component includes the effects associated with the limited size of the simulation sample and neglecting the data-based \mkee resolution correction. Two further sources of uncertainty are considered. First, rather than using the inclusive sample as discussed in \secref{sec:electron_fit}, an alternative parametrisation is derived using only the $\decay{ \Bp }{ K^{*+}(\to \Kp\piz)\epem}$ simulation sample. Second, to account for any uncertainty associated with the weighting of the $\decay{ \Bz }{ K^{*0}(\to \Kp\pim)\epem}$ and $\decay{ \Bp }{ K^{*+}(\to \Kp\piz)\epem}$ simulation samples to include a $K\pi$ $S$-wave contribution, an alternative model is obtained with this contribution doubled.

Two variations of the combinatorial model are considered, with a combined systematic uncertainty of $1.2\%$. As outlined in \secref{sec:electron_fit}, a correction using data to the combinatorial model is included by multiplying the phase-space model by a polynomial function. For the baseline model, the parameters of the polynomial function are determined by fitting the inverted BDT data, as shown on the left-hand side of \figref{fig:combinatorial_control}. The effect of using an alternative version of this correction is evaluated by determining its parameters by 
 instead fitting the \BKemu data. The combinatorial model additionally includes a correction that accounts for the effects of the BDT requirements. This correction is derived using the \BKemu sample, where the primary source of uncertainty stems from the limited size of the \BKemu sample. To address this, for each pseudoexperiment, an alternative combinatorial model is used that includes a BDT correction with parameters determined by fitting a bootstrapped version of the \BKemu sample.

There are multiple sources of systematic uncertainty related to the misidentified background component,  amounting to a combined uncertainty of 2.0\%. The first source of systematic uncertainty is associated with the choice to fix the parameters of the misidentified background model when fitting the rare electron data. This approach ignores significant uncertainties related to the model parameters, stemming primarily from the small size of the control samples utilised for their estimation. To address this, the parameters of the misidentified background are fluctuated for each determination of $R_K^{\rm{alternative}}$. The fluctuated parameter values are obtained by bootstrapping the control samples and repeating the data-based estimation method. A second source of systematic uncertainty involves varying the choice of model used to describe the misidentified background. As a variation to the baseline misidentified background model presented in \secref{sec:electron_fit}, a single KDE constructed using the weighted inverted PID data is used instead. Additional systematic uncertainties associated with the misidentified background component are evaluated by modifying elements of data control samples, such as changing the inverted PID requirements. 

Two systematic uncertainties are evaluated to account for the effects of backgrounds with yields considered too small to include in the baseline fit. One uncertainty pertains to backgrounds that include a \D meson, while the other relates to backgrounds that include a \psitwos resonance. Together, these uncertainties have a combined size of 0.6\%. Each systematic uncertainty is evaluated by fitting pseudoexperiments twice: first, using the baseline model, and second, by additionally injecting a background component. Each background is modelled with a KDE derived using simulation samples. The yield of each background component is fixed using selection efficiencies from simulation and branching fractions from Ref.~\cite{PDG2024}.

The newly adapted \RK extraction method significantly reduces any \qsqTrue model-dependence of this analysis. However, there remains a residual level of \qsqTrue model-dependence for the reasons discussed in \secref{sec:electron_fit}. Consequently, three sources of systematic uncertainty associated with the \qsqTrue model-independent extraction method are evaluated, which together contribute a total uncertainty of 0.6\%. Each source of systematic uncertainty is evaluated by varying an aspect of the approach and assessing the percentage shift in the parameter $C$, which by virtue of \equref{eq:RK_model_independent}, translates to an equivalent shift in \RK. The \qsqTrue model-independent approach implicitly assumes $\qsqTrue \equiv \qsq$ for muons. However, this assumption is not exact, so the effect of degrading the \qsq resolution in the muon mode samples is evaluated. The effect of explicitly varying the \qsqTrue model of simulated events is also evaluated. The baseline approach for computing $C$ uses simulated samples generated with a \qsqTrue model that incorporates resonances and interference effects. As a variation, the simplified \qsqTrue model, including only penguin contributions, is used. The final source of systematic uncertainty is associated with the binning of the $w_{\varepsilon}(\qsqTrue)$ weights, which is assessed by increasing the number of bins from the baseline value of 60 to 140.

Differences between the rare and resonant modes and imperfections in the applied corrections result in residual uncertainties related to the efficiencies. These uncertainties are computed by measuring shifts in \RK when alternative corrections are applied to the simulation. The total efficiency-related systematic uncertainty on \RK, including all contributions outlined in the following text, is $0.9\%$. In the baseline approach, the \BKJPsimumu data is used to derive corrections to the \Bu kinematic properties in the simulation thanks to the abundance and good resolution of that channel. Alternative corrections are derived by changing the variables corrected in the simulation and by using the \BKJPsiee data. Variations to the trigger corrections are also considered. Any potential bias introduced by the choice of tag is covered by deriving an alternative set of corrections, where the electron trigger response is computed on candidates that were instead triggered by the kaon candidate in the \BKJPsiee final state. Similar variations are studied by changing the trigger requirement applied to the control samples used to derive the PID corrections and by varying the kinematic binning used to obtain them. The systematic uncertainty associated with the mismodelling of particle multiplicity in simulation is determined by comparing the effect of adding to the initial sets of corrections three different multiplicity proxies.  Finally, the impact of the data-simulation corrections to the \qsqTrack resolution is covered by computing the variation in \RK when the correction is not applied. This systematic uncertainty is significantly reduced by the choice to select the \BKJPsiee mode with a \qsqTrack selection.

\section{Results and conclusions}\label{sec:results}
\begin{figure}[tb]
  \begin{center}
    \centering
    \includegraphics[page=1]{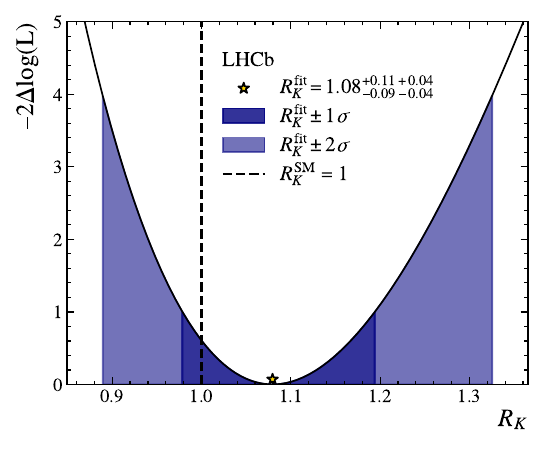}
    \vspace*{-0.5cm}
  \end{center}
  \caption{
    \small Likelihood (L) scan for \RK with the best-fit value (golden star) and expected SM value (vertical dashed line) highlighted. The $1\sigma$ and $2\sigma$ confidence regions are indicated by the dark and light-shaded regions, respectively.}
  \label{fig:profile}
\end{figure}

The measured value for \RK at high-\qsq, corrected for bias intrinsic to the rare electron mode fit, is 
\begin{center}
    $\RK(\qsq>14.3$\;GeV$^2/c^4) = 1.08^{+0.11\;+0.04}_{-0.09\;-0.04}$\;,
\end{center}
where the first uncertainty is statistical and the second is systematic. The data used for the measurement consists of beauty-meson decays produced in proton-proton collisions, corresponding to an integrated luminosity of $9\invfb$, collected by the \lhcb experiment between 2011 and 2018. Due to the limited sample size of the rare electron mode data, the statistical uncertainty of \RK is not Gaussian. Therefore, the confidence intervals of \RK are obtained through the likelihood profile shown in \figref{fig:profile}.

To determine separate statistical and systematic uncertainties, the likelihood profile is repeated twice, once excluding the contribution from systematic effects in the likelihood function and again including them. The former profile returns the statistical uncertainty, whereas the latter profile yields the total uncertainty on \RK. The systematic uncertainty is then determined by subtracting the statistical from the total uncertainty in quadrature.

This is the first measurement of \RK in the high-\qsq region at a hadron collider and the result is compatible with the SM and with previous measurements~\cite{belle_rk}. The measurement has complementary dependence on both background and efficiency mismodelling to measurements below the \psitwos resonance~\cite{LHCb-PAPER-2022-045,LHCb-PAPER-2022-046} and is the most precise measurement of LU in the kinematic region above the \psitwos resonance~\cite{belle_rk,LHCb-PAPER-2024-032}.

\section*{Acknowledgements}
%
%
\noindent We express our gratitude to our colleagues in the CERN
accelerator departments for the excellent performance of the LHC. We
thank the technical and administrative staff at the LHCb
institutes.
We acknowledge support from CERN and from the national agencies:
ARC (Australia);
CAPES, CNPq, FAPERJ and FINEP (Brazil); 
MOST and NSFC (China); 
CNRS/IN2P3 (France); 
BMBF, DFG and MPG (Germany); 
INFN (Italy); 
NWO (Netherlands); 
MNiSW and NCN (Poland); 
MCID/IFA (Romania); 
MICIU and AEI (Spain);
SNSF and SER (Switzerland); 
NASU (Ukraine); 
STFC (United Kingdom); 
DOE NP and NSF (USA).
We acknowledge the computing resources that are provided by ARDC (Australia), 
CBPF (Brazil),
CERN, 
IHEP and LZU (China),
IN2P3 (France), 
KIT and DESY (Germany), 
INFN (Italy), 
SURF (Netherlands),
Polish WLCG (Poland),
IFIN-HH (Romania), 
PIC (Spain), CSCS (Switzerland), 
and GridPP (United Kingdom).
We are indebted to the communities behind the multiple open-source
software packages on which we depend.
Individual groups or members have received support from
Key Research Program of Frontier Sciences of CAS, CAS PIFI, CAS CCEPP, 
Fundamental Research Funds for the Central Universities,  and Sci.\ \& Tech.\ Program of Guangzhou (China);
Minciencias (Colombia);
EPLANET, Marie Sk\l{}odowska-Curie Actions, ERC and NextGenerationEU (European Union);
A*MIDEX, ANR, IPhU and Labex P2IO, and R\'{e}gion Auvergne-Rh\^{o}ne-Alpes (France);
Alexander-von-Humboldt Foundation (Germany);
ICSC (Italy); 
Severo Ochoa and Mar\'ia de Maeztu Units of Excellence, GVA, XuntaGal, GENCAT, InTalent-Inditex and Prog.~Atracci\'on Talento CM (Spain);
SRC (Sweden);
the Leverhulme Trust, the Royal Society and UKRI (United Kingdom).

\addcontentsline{toc}{section}{References}
\bibliographystyle{LHCb}
\bibliography{main,standard,LHCb-PAPER,LHCb-CONF,LHCb-DP,LHCb-TDR}

\newpage
\centerline
{\large\bf LHCb collaboration}
\begin
{flushleft}
\small
R.~Aaij$^{38}$\lhcborcid{0000-0003-0533-1952},
A.S.W.~Abdelmotteleb$^{57}$\lhcborcid{0000-0001-7905-0542},
C.~Abellan~Beteta$^{51}$\lhcborcid{0009-0009-0869-6798},
F.~Abudin{\'e}n$^{57}$\lhcborcid{0000-0002-6737-3528},
T.~Ackernley$^{61}$\lhcborcid{0000-0002-5951-3498},
A. A. ~Adefisoye$^{69}$\lhcborcid{0000-0003-2448-1550},
B.~Adeva$^{47}$\lhcborcid{0000-0001-9756-3712},
M.~Adinolfi$^{55}$\lhcborcid{0000-0002-1326-1264},
P.~Adlarson$^{83}$\lhcborcid{0000-0001-6280-3851},
C.~Agapopoulou$^{14}$\lhcborcid{0000-0002-2368-0147}
% ...

{\footnotesize \it

% ...
$^{14}$Universit{\'e} Paris-Saclay, CNRS/IN2P3, IJCLab, Orsay, France\\
% ...
$^{38}$Nikhef National Institute for Subatomic Physics, Amsterdam, Netherlands\\
% ...
$^{47}$Instituto Galego de F{\'\i}sica de Altas Enerx{\'\i}as (IGFAE), Universidade de Santiago de Compostela, Santiago de Compostela, Spain\\
% ...
$^{51}$Physik-Institut, Universit{\"a}t Z{\"u}rich, Z{\"u}rich, Switzerland\\
% ...
$^{55}$H.H. Wills Physics Laboratory, University of Bristol, Bristol, United Kingdom\\
% ...
$^{57}$Department of Physics, University of Warwick, Coventry, United Kingdom\\
% ...
$^{60}$School of Physics and Astronomy, University of Glasgow, Glasgow, United Kingdom\\
$^{61}$Oliver Lodge Laboratory, University of Liverpool, Liverpool, United Kingdom\\
% ...
$^{69}$Syracuse University, Syracuse, NY, United States\\
% ...
$^{83}$Department of Physics and Astronomy, Uppsala University, Uppsala, Sweden, associated to $^{60}$\\
% ...
}
\end{flushleft}

\end{document}